\documentclass[twocolumn,english,aps,manuscript,prb,floatfix,superscriptaddress,showpacs,amsmath,amssymb,reprint]{revtex4-1}
\usepackage[T1]{fontenc}
\usepackage[latin9]{inputenc}
\usepackage{color}
\usepackage{babel}
\usepackage{amsmath}
\usepackage{graphicx}
\usepackage{esint}
\usepackage[unicode=true,pdfusetitle,
 bookmarks=true,bookmarksnumbered=false,bookmarksopen=false,
 breaklinks=true,pdfborder={0 0 0},backref=false,colorlinks=true]
 {hyperref}
\hypersetup{
 linkcolor=blue, citecolor=blue, urlcolor=blue}

\makeatletter
\@ifundefined{textcolor}{}
{%
 \definecolor{BLACK}{gray}{0}
 \definecolor{WHITE}{gray}{1}
 \definecolor{RED}{rgb}{1,0,0}
 \definecolor{GREEN}{rgb}{0,1,0}
 \definecolor{BLUE}{rgb}{0,0,1}
 \definecolor{CYAN}{cmyk}{1,0,0,0}
 \definecolor{MAGENTA}{cmyk}{0,1,0,0}
 \definecolor{YELLOW}{cmyk}{0,0,1,0}
}

\makeatother

\begin{document}

\title{Strong-disorder renormalization-group study of the one-dimensional
tight-binding model}

\author{H. Javan Mard}

\affiliation{Department of Physics and National High Magnetic Field Laboratory,
Florida State University, Tallahassee, FL 32306}

\author{José A. Hoyos}

\affiliation{Instituto de Física de São Carlos, Universidade de São Paulo, C.P.
369, São Carlos, SP 13560-970, Brazil}

\author{E. Miranda}

\affiliation{Instituto de Física Gleb Wataghin, Unicamp, R. Sérgio Buarque de
Holanda, 777, Campinas, SP 13083-859, Brazil}

\author{V. Dobrosavljevi\'{c}}

\affiliation{Department of Physics and National High Magnetic Field Laboratory,
Florida State University, Tallahassee, FL 32306}
\begin{abstract}
We formulate a strong-disorder renormalization-group (SDRG) approach
to study the beta function of the tight-binding model in one dimension
with both diagonal and off-diagonal disorder for states at the band
center. We show that the SDRG method, when used to compute transport
properties, yields exact results since it is identical to the transfer
matrix method. The beta function is shown to be universal when only
off-diagonal disorder is present even though single-parameter scaling
is known to be violated. A different single-parameter scaling theory
is formulated for this particular (particle-hole symmetric) case.
Upon breaking particle-hole symmetry (by adding diagonal disorder),
the beta function is shown to crossover from the universal behavior
of the particle-hole symmetric case to the conventional non-universal
one in agreement with the two-parameter scaling theory. We finally
draw an analogy with the random transverse-field Ising chain in the
paramagnetic phase. The particle-hole symmetric case corresponds to
the critical point of the quantum Ising model while the generic case
corresponds to the Griffiths paramagnetic phase.
\end{abstract}
\maketitle

\section{Introduction}

The tight-binding model in one dimension is one of the most studied
models for Anderson localization. It is well established that away
from the band center and band edges,\ \cite{deych-lisyansky-altshuler-prl00,deych-lisyansky-altshuler-prb01,deych-etal-prl03}
single-parameter scaling theory holds and predicts a universal beta
function for the conductance.\ \cite{abrahams-gang4,anderson-etal-prb80,cohen-etal-prb88}
Near the band center, however, a two-parameter scaling theory is required.\ \cite{balents-fisher-prb97}

Generically at the band center, violations to the single-parameter
scaling theory are known to be small\ \cite{schomerus-titov-prb03}
and thus, tiny deviations from the universal beta function are expected.
On the other hand, when particle-hole symmetry is present (off-diagonal
disorder only), strong deviations are expected since a ``delocalization''
transition takes place.\ \cite{eggarter-riedinger-prb78,soukoulis-economou-prb81,inui-trugman-abrahams-prb94,balents-fisher-prb97}
However, universality is somehow recovered.\ \cite{soukoulis-economou-prb81}

It is easy to see that single-parameter scaling leads to a universal
beta function. In this case, scaling implies that the average conductance
$g$ (in one dimension, the geometric average~\cite{anderson-etal-prb80})
depends on the disorder strength $W$ and the system size $L$ only
through the combination $L/\xi\left(W\right)$, $g=g\left(L/\xi\right)$,
where $\xi\left(W\right)$, the relevant parameter, is the localization/correlation
length. It follows immediately that the beta function, when expressed
in terms of $g$, is a universal function 
\begin{eqnarray}
\beta\left(g\right) & = & d\ln g/d\ln L=\frac{\left(L/\xi\right)g^{\prime}\left(L/\xi\right)}{g\left(L/\xi\right)}\nonumber \\
 & = & \frac{g^{-1}\left(g\right)g^{\prime}\left[g^{-1}\left(g\right)\right]}{g}.\label{eq:betauniversal}
\end{eqnarray}
If, on the other hand, a second parameter $c$ is required, $g=g\left(L/\xi,c\right)$,
then, in general, $\beta\left(g\right)$ will also depend on $c$
and be non-universal.

These considerations have a clear signature when we consider the full
distribution of sample conductances $g_{s}$ for a given system size
$L$. In general, the number of independent parameters required for
the description of the distribution (its various cumulants, for example)
determines the corresponding scaling behavior. It should be noted,
however, that most discussions focus on the large-$L$ limit of this
distribution only. Then, in one dimension, even if two-parameter scaling
holds, $g\sim\exp\left(-L/\xi\right)$ and the beta function is universal,
$\beta\left(g\right)\approx\ln g$. The distribution is log-normal
and the second parameter only affects the variance.\cite{schomerus-titov-prb03}
Thus, the non-universality of the beta function is only seen at next-to-leading
order in the large-$L$ limit.

A similar dichotomy is encountered at certain disordered critical
points governed by infinite-randomness fixed points (which are universal)
surrounded by quantum Griffiths phases (which are not)\ \cite{fisher95}.
In this case, a suitable description via effective distributions of
the system couplings was possible due to a strong-disorder renormalization
group (SDRG) method.\ \cite{MDH-PRL,MDH-PRB,bhatt-lee} This suggests
that this method might be specially suitable for the study of the
universality properties of the tight-binding model.

Here, we revisit this model (with diagonal and off-diagonal disorder)
by focusing on the transport properties of the band center state.
We explicitly investigate the universal (non-universal) behavior of
the beta function when particle-hole symmetry is present (broken)
using numerically and analytically exact methods as well as the SDRG
method. It is shown that the latter is equivalent to the transfer
matrix method in one dimension and thus, gives exact results. Moreover,
its advantage is in its simplicity which allows us to compute the
beta function in a straightforward manner. We confirm that, for the
particle-hole symmetric case, the distribution of (the properly scaled)
conductance is universal and thus, a different single-parameter scaling
theory applies. This difference stems from the fact that the state
is stretched-exponentially localized, in contrast to the usual exponentially
localized states.

The remaining of this article is organized as follows. In Sec.\ \ref{sec:Model-SDRG},
we define the model and derive the SDRG transformations. In Sec.\ \ref{sec:The-beta-function},
we discuss the computation of the beta function in general. Then,
the particle-hole symmetric case is discussed in Sec.~\ref{sec:PH-symmetry}
and the generic one in Sec.~\ref{sec:The-generic-case}. In order
to make connection to single- and two-parameter scaling theories,
we briefly analyze the Lyapunov exponent in Sec.\ \ref{sec:The-Lyapunov-exponent}.
Finally, we leave our conclusions and final remarks to Sec.\ \ref{sec:Conclusion}.

\section{The model and the SDRG method\label{sec:Model-SDRG}}

Consider the one-dimensional tight-binding model

\begin{equation}
H=\sum_{i}\left[\varepsilon_{i}c_{i}^{\dagger}c_{i}^{\phantom{\dagger}}+t_{i,i+1}\left(c_{i}^{\dagger}c_{i+1}^{\phantom{\dagger}}+{\rm h.c.}\right)\right],\label{eq:Hamiltonian}
\end{equation}
where $c_{i}^{\dagger}$($c_{i}^{\phantom{\dagger}}$) is the canonical
creation (annihilation) operator of spinless fermions at site $i$,
$t_{i,j}=t_{j,i}=t_{i}\delta_{j,i+1}$ is the hopping amplitude between
nearest-neighbor sites and $\varepsilon_{i}$ is the onsite energy.
Both diagonal and off-diagonal energies are independent random variables
drawn from arbitrary initial distributions. This model has been thoroughly
studied\ \cite{50-years-localization} but still continues to present
surprises.\ \cite{johri-bhatt-prl12,johri-bhatt-prb12} It is known
that any amount of disorder renders all states exponentially localized,\ \textcolor{red}{\cite{mott-twose-advphys61}}
except for the case of off-diagonal disorder only ($\varepsilon_{i}=0$),
in which the middle-band state is stretched-exponentially localized.\ \cite{eggarter-riedinger-prb78,inui-trugman-abrahams-prb94}

In order to treat this model using the SDRG philosophy,\ \cite{MDH-PRL,MDH-PRB,bhatt-lee}
we first locate the largest energy constant in the Hamiltonian and
identify it as the cutoff of our problem, i.e., $\Omega=\max\left\{ |t_{i}|,\,|\varepsilon_{i}|\right\} $. 

\begin{figure}[h]
\begin{centering}
\includegraphics[width=0.9\columnwidth]{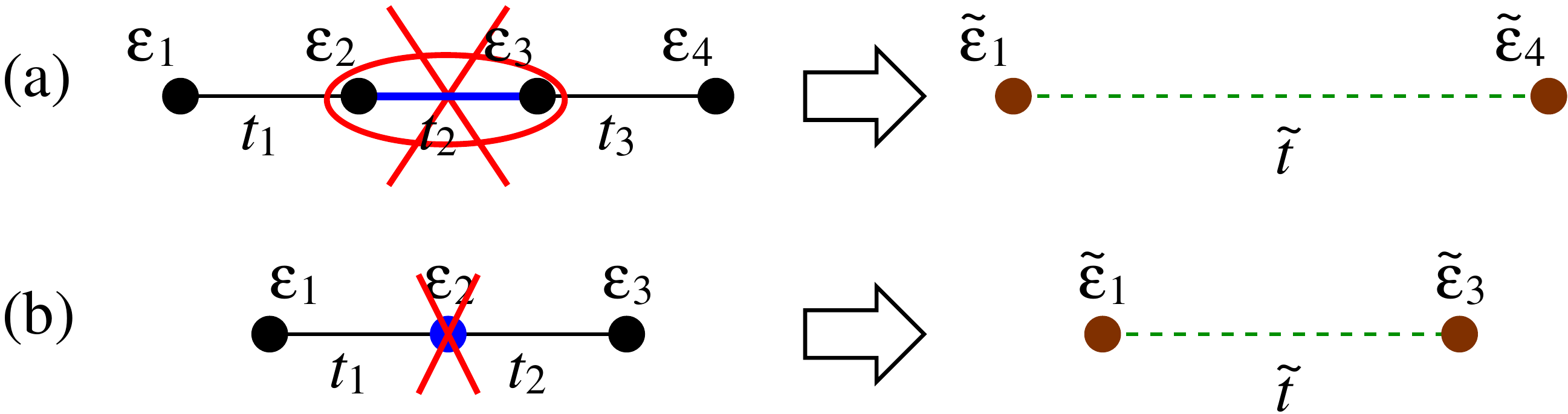}
\par\end{centering}

\protect\caption{\label{fig:decimation}Schematic decimation procedure for (a) bond,
and (b) site transformations.}
\end{figure}

Consider the case in which the hopping term happens to be the largest
energy scale, say $\Omega=|t_{2}|$ {[}see Fig.\ \hyperref[fig:decimation]{\ref{fig:decimation}(a)}{]}.
The resonant and anti-resonant states between sites 2 and 3 lie at
the top and the bottom of the band. Since we are interested in analyzing
the band-center state, both states are then discarded and only the
virtual tunneling process between sites 1 and 4 is kept. The renormalized
hopping and onsite energies then read (see App.\ \ref{sec:SDRG})

\begin{eqnarray}
\tilde{t}_{1,4} & = & -t_{1}t_{2}t_{3}/\left(t_{2}^{2}-\varepsilon_{2}\varepsilon_{3}\right),\label{eq:t-tdec}\\
\tilde{\varepsilon}_{j} & = & \varepsilon_{j}+\left(\varepsilon_{3}t_{1}^{2}\delta_{j,1}+\varepsilon_{2}t_{3}^{2}\delta_{j,4}\right)/\left(t_{2}^{2}-\varepsilon_{2}\varepsilon_{3}\right).\label{eq:e-tdec}
\end{eqnarray}
On the other hand, if the strongest energy scale is an onsite energy,
say $\Omega=|\varepsilon_{2}|$ {[}see Fig.\ \hyperref[fig:decimation]{\ref{fig:decimation}(b)}{]},
then the particle would be localized at or repelled from site 2 depending
on the sign of $\varepsilon_{2}$. Again, this corresponds to states
away from the band center and thus, site 2 is removed from the chain.
The renormalized couplings then read (see App.\ \ref{sec:SDRG}) 

\begin{eqnarray}
\tilde{t}_{1,3} & = & -t_{1}t_{2}/\varepsilon_{2},\label{eq:t-edec}\\
\tilde{\varepsilon}_{i} & = & \varepsilon_{i}-t_{2,i}^{2}/\varepsilon_{2}.\label{eq:e-edec}
\end{eqnarray}

We report that the SDRG transformations Eqs.\ (\ref{eq:t-tdec})---(\ref{eq:e-edec}),
although computed in perturbation theory, are \emph{exact} transformations
as long as one is interested in transport properties (transmittance)
only (see App.\ \ref{sec:Transfer-matrix}). As a consequence, the
SDRG method yields exact results for the beta function in one dimension.
Finally, we remark that these transformations recover the ones in
the literature in the appropriate limit of approximation.\ \cite{motrunich-damle-huse-prb02,melin-doucot-igloi-prb05}

Given that these transformations are exact, there is no need to either
search for the largest energy scale in the system or to introduce
the cutoff $\Omega$. One can iterate Eqs.\ (\ref{eq:t-tdec})---(\ref{eq:e-edec})
\emph{in arbitrary order} until all sites are decimated, leaving the
effective trio/link $\tilde{\varepsilon}_{1}$, $\tilde{t}_{1,L}$,
and $\tilde{\varepsilon}_{L}$ connected to external leads: the conductance
is then easily computed. However, we keep the SDRG formulation because
it allows us to perform an analytical treatment in one dimension,
as we show later. Furthermore, the main purpose of using the SDRG
formulation is that it can be applied in higher dimensions and/or
in the presence of interactions. In these cases, the SDRG transformation
is no longer exact and the hierarchical decimation scheme is needed
to correctly justify the perturbation theory.

\section{The beta function\label{sec:The-beta-function}}

In the following Sections, we compute the beta function using analytical
results from the SDRG method and compare with exact results. For such
a task, we use the dimensionless conductance defined as

\begin{equation}
g=\left[T/\left(1-T\right)\right]_{{\rm geo}}=\exp\left\langle \ln\left[T/\left(1-T\right)\right]\right\rangle \label{eq:g-average}
\end{equation}
where $T$ is the transmittance, $\left\langle \cdots\right\rangle $
means the disorder average and $\left(\cdots\right)_{{\rm geo}}$
denotes the geometric average, which we use for the typical value.
It should be noticed that one may use different definitions of $g$
such as $\left(T\right)_{{\rm geo}}$ or $\frac{\left(T\right)_{{\rm geo}}}{1-\left(T\right)_{{\rm geo}}}$
. Subtleties about these definitions are not important here (see more
details in App.\ \ref{sec:Alternative-conductance}). The transmittance
is computed using the effective trio $\tilde{\varepsilon}_{1}$, $\tilde{t}_{1,L}$,
and $\tilde{\varepsilon}_{L}$ (for which we drop the tildes henceforth):
\begin{equation}
T=\frac{\left(2t_{1,L}t_{0}\right)^{2}}{\left(t_{0}^{2}+t_{1,L}^{2}+\varepsilon_{1}\varepsilon_{L}\right)^{2}+t_{0}^{2}\left(\varepsilon_{1}-\varepsilon_{L}\right)^{2}},\label{eq:transmitance}
\end{equation}
where $t_{0}$ is the hopping constant of the leads. In order to have
a well-defined Ohmic regime, we need to set $t_{0}=\Omega_{0}$, the
bare energy cutoff of the distributions of $t$'s and $\varepsilon$'s.
The beta function is then computed via 
\begin{equation}
\beta=\frac{{\rm d}\ln g}{{\rm d}\ln L},\label{eq:beta-function}
\end{equation}
where $L$ is the system size.

The strategy from now on is to compute the typical value of $g$ (and
thus, $\beta$) using the effective probability for $\varepsilon_{1}$,
$t_{1,L}$, and $\varepsilon_{L}$ given by the SDRG method. Analytical
results are not simple and limited. Therefore, we compare with exact
results obtained either by another analytical method or by numerical
implementation of the transformations in Eqs.\ (\ref{eq:t-tdec})---(\ref{eq:e-edec}).

\section{The particle-hole symmetric case\label{sec:PH-symmetry}}

In this section, we compute the beta function for the case in which
all onsite energies are zero ($\varepsilon_{i}=0$) in the Hamiltonian
of Eq.\ (\ref{eq:Hamiltonian}) in different approaches and compare
them.

\subsection{Analytical SDRG results\label{sub:PH-SDRG}}

In this simpler case, only the transformation in Eq.\ (\ref{eq:t-tdec})
is needed. Notice that, except for an irrelevant sign, the SDRG decimation
procedure is identical to that of the random transverse field Ising
model \emph{at} \emph{criticality}\ \cite{fisher95} as could be
anticipated by a Wigner-Jordan mapping between these two models. Moreover,
the transmittance simplifies to 
\begin{equation}
T=\left(2t_{1,L}/\Omega_{0}\right)^{2}\left[1+\left(t_{1,L}/\Omega_{0}\right)^{2}\right]^{-2}.\label{eq:transmitancePH}
\end{equation}

Running down the energy scale $\Omega$, the fixed-point distribution
for the hoppings is\ \cite{fisher95}

\begin{equation}
P(t)=\Theta(\Omega-\left|t\right|)\frac{1}{\Omega u(\Omega)}\left(\frac{\Omega}{\left|t\right|}\right)^{1-1/u(\Omega)},\label{eq:PH-distribtion}
\end{equation}
where $\Theta\left(x\right)$ is the Heaviside function, $u(\Omega)=u_{0}+\Gamma$
is a slowly varying exponent with $\Gamma=\ln\left(\Omega_{0}/\Omega\right)$,
$\Omega_{0}$ is the cutoff of the bare distribution of $t$'s, and
$u_{0}$ is proportional to the disorder strength of the bare system.
This fixed-point distribution is universal in the sense that it attracts
all initial distributions\ %
\footnote{Except for extremely singular ones like $P\sim1/\left[\left|t\right|\left|\ln\left|t\right|\right|^{x}\right]$.%
} as the limit $\Omega\rightarrow0$ is approached. Moreover, since
the system disorder increases along the RG flow ($\left\langle t^{2}\right\rangle /\left\langle t\right\rangle ^{2}\rightarrow\infty$
as $\Omega\rightarrow0$), the associated fixed point is of infinite
randomness type.

In order to compute $g$ and $\beta$, we need the distribution of
$t_{1,L}$ and its dependence on the system size $L$. Using the results
of Refs.\ \onlinecite{fisher-young-RTFIM,hoyosvieiralaflorenciemiranda}
in the limit $L\gg1$, the distribution of the last hopping is 
\begin{align}
{\cal P}\left(\eta\right) & =\frac{4}{\sqrt{\pi}}\sum_{n=0}^{\infty}\left(-1\right)^{n}\left(n+\frac{1}{2}\right)e^{-\eta^{2}\left(n+\frac{1}{2}\right)^{2}},\label{eq:P(eta)}\\
 & =\frac{4\pi}{\eta^{3}}\sum_{n=0}^{\infty}\left(-1\right)^{n}\left(n+\frac{1}{2}\right)e^{-\pi^{2}\left(n+\frac{1}{2}\right)^{2}/\eta^{2}},\label{eq:P(eta)2}
\end{align}
where $\eta=\ln\left(\Omega_{0}/t_{1,L}\right)/\left(u_{0}\sqrt{L/2}\right)$.
Note that the sample conductance can be written as
\begin{equation}
g_{s}=\frac{1}{\sinh^{2}\zeta_{1,L}},\label{eq:sampleg}
\end{equation}
where $\zeta_{1,L}=\ln\left(\Omega_{0}/t_{1,L}\right)$, which, through
${\cal P}\left(\eta\right)$, yields the distribution of $g_{s}$.

\begin{figure}[t]
\begin{centering}
\includegraphics[clip,width=1\columnwidth]{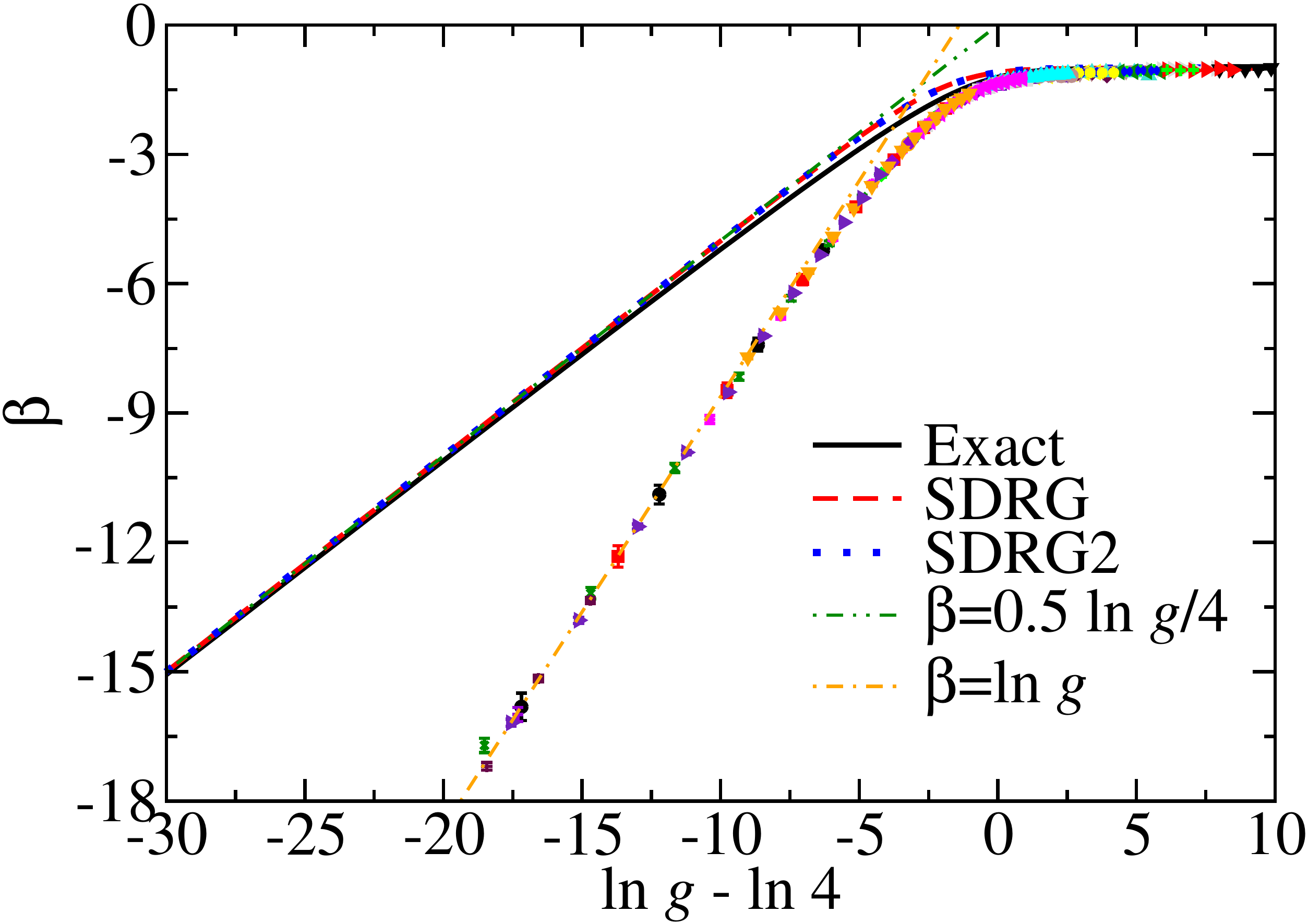}
\par\end{centering}

\protect\caption{The beta function for the particle-hole symmetric case ($\varepsilon_{i}=0$,
continuous, dashed and dotted lines) and for the more conventional
case with diagonal disorder only ($t_{i}=t_{0}$, symbols). The SDRG
(dashed red line), exact (continuous black line), and SDRG2 (dotted
blue line) results are discussed in Secs.\ \ref{sub:PH-SDRG}, \ref{sub:PH-Exact},
and \ref{sub:PH-simpleSDRG}, respectively. \label{fig:The-beta-function}}
\end{figure}

The conductance and the beta function are thus 

\begin{eqnarray}
\ln g & = & \ln4-\alpha\sqrt{\pi}-2\left\langle \ln\left(1-e^{-\eta\alpha}\right)\right\rangle ,\label{eq:lngPH}\\
\beta & = & -\alpha\sqrt{\pi}/2-\left\langle \eta\alpha/\left(e^{\eta\alpha}-1\right)\right\rangle ,\label{eq:betafxnPH}
\end{eqnarray}
where $\alpha=u_{0}\sqrt{2L}$ and we used that $\left\langle \eta\right\rangle =\sqrt{\pi}$.
Notice that Eqs.\ (\ref{eq:lngPH}) and (\ref{eq:betafxnPH}) give
$\beta$ as a function of $\ln g$ parameterized by $\alpha$. Thus,
the beta function is universal as expected from the universality of
${\cal P}\left(\eta\right)$. A simple numerical integration of these
equations is shown in Fig.\ \ref{fig:The-beta-function} as a dashed
red line.

The Ohmic regime is easily accessed in the limit $\alpha\rightarrow0$.
Expanding $e^{\eta\alpha}$ in powers of $\alpha$ and using Eq.\ (\ref{eq:P(eta)})
for the averaging, we find $\left\langle \eta\alpha/\left(e^{\eta\alpha}-1\right)\right\rangle \approx1-\sqrt{\pi}\alpha/2+G\alpha^{2}/3$,
where $G\approx0.916$ is the Catalan constant. Similarly, $\left\langle \ln\left(1-e^{-\eta\alpha}\right)\right\rangle \approx\ln\alpha-\left(\gamma+\ln4\right)/2+2X-\alpha\sqrt{\pi}/2$,
where $\gamma\approx0.577$ is Euler's constant and $X=-\sum_{n=0}^{\infty}\left(-1\right)^{n}\ln\left(n+1/2\right)\approx0.738$.
Thus, $\beta\approx-1-Y/g$, where $Y=16Ge^{\gamma-4X}/3\approx0.454$.

The localized regime is obtained straightforwardly in the limit $\alpha\rightarrow\infty$.
Deep in the localized regime, $\ln g\rightarrow-\alpha\sqrt{\pi}$
and $\beta\rightarrow-\alpha\sqrt{\pi}/2$, and thus, $\beta=\frac{1}{2}\ln g$.
Notice that this is not the usual localized regime behavior\ \cite{abrahams-gang4},
for which $\beta=\ln g$. The $\frac{1}{2}$ factor can be understand
as follows. For the particle-hole symmetric case, the wave function
is stretched-exponentially localized:\ \cite{soukoulis-economou-prb81,inui-trugman-abrahams-prb94}
$\left|\psi(r)\right|^{2}\sim e^{-\sqrt{r}}$, where $r$ is the distance
from the central site in units of the associated localization length.
The transmittance of a chain of size $L$ much larger than the localization
length is thus $\ln g\approx\left\langle \ln T\right\rangle \approx\ln\left|\psi\right|^{2}\sim-\sqrt{L}$.
Therefore, $\beta=\frac{1}{2}\ln g$.

Obtaining corrections to the strongly localized regime requires tedious
algebra. Using Eq.\ (\ref{eq:P(eta)2}), the mean values can be obtained
in the saddle-point approximation. In addition, we keep only the dominant
term $n=0$. Then, $-2\left\langle \ln\left(1-e^{-\eta\alpha}\right)\right\rangle \approx2\left\langle e^{-\eta\alpha}\right\rangle =2\sqrt{\pi}\left(3/2\right)^{4}A^{-5/2}\alpha^{-8/3}e^{-A\alpha^{2/3}}$,
with $A=3\pi^{2/3}/2^{4/3}$ and $\left\langle \eta\alpha/\left(e^{\eta\alpha}-1\right)\right\rangle \approx\left\langle \eta\alpha e^{-\eta\alpha}\right\rangle =\sqrt{\pi}\left(3/2\right)^{3}A^{-3/2}\alpha^{-2}e^{-A\alpha^{2/3}}$.
Finally, 
\begin{equation}
\beta\approx\frac{1}{2}\ln\left(g/4\right)-3\sqrt{3\pi}\frac{e^{-A\pi^{-1/3}\left(-\ln g/4\right)^{2/3}}}{2\ln^{2}g}.\label{eq:bfuncsdrg}
\end{equation}

\subsection{Exact results\label{sub:PH-Exact}}

It is interesting to compare the above analytical SDRG results with
exact ones. As pointed out, the SDRG decimations rules Eqs.\ (\ref{eq:t-tdec})---(\ref{eq:e-edec})
are exact if we are interested in computing transport properties.
Therefore, for the particle-hole symmetric case, the last hopping
constant can be easily computed as 
\begin{equation}
t_{1,L}=\frac{t_{1}t_{3}\dots t_{L-1}}{t_{2}t_{4}\dots t_{L-2}},\label{eq:t-Exact}
\end{equation}
where we are considering chains with an even number $L$ sites attached
to leads and neglecting the unimportant negative sign in Eq.\ (\ref{eq:t-tdec}).
Defining $\zeta_{i}=\ln\left(\Omega_{0}/t_{i}\right)$, we find that
$\zeta_{1,L}$ is the result of a random walk in the $\zeta_{i}$
space. In the $L\gg1$ limit, the central limit theorem can be used
to find the distribution of $\zeta_{1,L}=\ln\left(\Omega_{0}/t_{1,L}\right)$:
$Q\left(\zeta_{1,L}\right)=e^{-\left(\zeta_{1,L}-\left\langle \zeta\right\rangle _{0}\right)^{2}/\left(2\sigma^{2}\right)}/\left(\sqrt{2\pi}\sigma\right),$
where $\sigma^{2}=\left(L-1\right)\sigma_{0}^{2}$, and $\left\langle \zeta\right\rangle _{0}$
and $\sigma_{0}^{2}$ are the mean and the variance of the bare distribution
of $\zeta_{i}=\ln\left(\Omega_{0}/t_{i}\right)$, respectively. Notice
that universality is obtained in the limit of small disorder $\left\langle \zeta\right\rangle _{0}\rightarrow0$
and large system size $L\rightarrow\infty$. As we want to compare
this approach with the analytical SDRG one, we define $\eta=\zeta_{1,L}/\left(\sigma_{0}\sqrt{L/2}\right)$
and find 
\begin{equation}
{\cal Q}\left(\eta\right)=\frac{1}{2\sqrt{\pi}}e^{-\frac{1}{4}\eta^{2}},\label{eq:Q(eta)}
\end{equation}
which recovers the results of Ref.\ \onlinecite{soukoulis-economou-prb81}.
Using Eq.~(\ref{eq:sampleg}), the distribution of conductance samples
is easily obtained. In particular, when $L\to\infty$, $\ln g_{s}\approx-2\left|\zeta_{1,L}\right|$
and the distribution of $g_{s}$ is log-normal. The conductance and
the beta function are given, respectively, by

\begin{eqnarray}
\ln g & = & \ln4-\left\langle \ln\left(1-e^{-\eta\alpha}\right)^{2}\right\rangle ,\label{eq:lng-Exact}\\
\beta & = & -\left\langle \eta\alpha/\left(e^{\eta\alpha}-1\right)\right\rangle ,\label{eq:beta-Exact}
\end{eqnarray}
where $\alpha=\sigma_{0}\sqrt{2L}$ and $\left\langle \eta\right\rangle =0$.
Again, we have parameterized $\beta$ and $\ln g$ in terms of $\alpha$,
which implies a universal beta function. The exact beta function is
plotted in Fig.\ \ref{fig:The-beta-function} as a solid black line. 

For $\alpha\ll1$, we expand the averages in powers of $\alpha$.
Thus, $\left\langle \ln\left(1-e^{-\eta\alpha}\right)^{2}\right\rangle \approx\left\langle \ln\left(\alpha\eta\right)^{2}\right\rangle -{\cal O}\left(\alpha\right)^{2}=2\ln\alpha+\gamma$,
and $\left\langle \eta\alpha/\left(e^{\eta\alpha}-1\right)\right\rangle \approx1+\left\langle \left(\alpha\eta\right)^{2}\right\rangle /12+{\cal O}\left(\alpha\right)^{4}=1+\alpha^{2}/6$.
We then find that $\beta=-1-Y^{\prime}/g$, with $Y^{\prime}=2e^{\gamma}/3\approx1.187$.

The localized regime ($\alpha\rightarrow\infty$) is easily obtained
by noticing that the averages in Eqs.\ (\ref{eq:lng-Exact}) and
(\ref{eq:beta-Exact}) are dominated by the negative values of $\eta$.
Then, we simplify $\left\langle \ln\left(1-e^{-\eta\alpha}\right)^{2}\right\rangle \approx\left\langle -2\alpha\eta\Theta\left(-\eta\right)\right\rangle =2\alpha/\sqrt{\pi}$,
and $\left\langle \eta\alpha/\left(e^{\eta\alpha}-1\right)\right\rangle \approx-\left\langle \alpha\eta\right\rangle =\alpha/\sqrt{\pi}$.
Thus, $\beta=\frac{1}{2}\ln g$ as expected. 

Obtaining corrections to the strongly localized regime is not as simple.
We use that $\ln\left(1-e^{-x}\right)^{2}=-2[x\Theta\left(-x\right)+\sum_{n=1}^{\infty}e^{-n\left|x\right|}/n]$.
The resulting integrals are error functions ${\rm Eff}\left(\alpha n\right)$
which we further expand in the limit of large argument: ${\rm Eff}\left(x\right)=1-e^{-\left(x\right)^{2}}\left(x\sqrt{\pi}\right)^{-1}[1-1/(2x^{2})+\dots]$.
The final result is that $\left\langle \ln\left(1-e^{-\eta\alpha}\right)^{2}\right\rangle \approx2\alpha\pi^{-1/2}[1-\pi^{2}/(6\alpha^{2})]$,
and 
\begin{equation}
\beta\approx-\alpha\pi^{-1/2}[1+\pi^{2}/(6\alpha^{2})]\approx\frac{1}{2}\ln g-\ln2+\frac{2\pi}{3\ln g}.\label{eq:bfuncexact}
\end{equation}

\subsection{Comparison between analytical SDRG and exact results}

Let us compare the analytical SDRG results Eqs.\ (\ref{eq:P(eta)})
and (\ref{eq:P(eta)2}) with the exact ones Eq.\ (\ref{eq:Q(eta)}). 

The main difference is that $\eta$ is distributed only among positive
values in ${\cal P}\left(\eta\right)$ while it can assume both positive
and negative values in ${\cal Q}\left(\eta\right)$. This may seem
due to the hierarchical decimation procedure of the SDRG: the new
renormalized hopping is always less (in magnitude) than the decimated
ones {[}see Eq.\ (\ref{eq:t-tdec}){]}. Hence, $\eta$ in the SDRG
scheme is necessarily positive. However, remember that the SDRG transformation
in Eq.\ (\ref{eq:t-tdec}) is an exact one. The problem in the hierarchical
scheme is the inability of handling the boundary conditions correctly.
To be precise, consider the simple case of a 4-site long chain. The
exact effective hopping between sites 1 and 4, given by Eq.\ (\ref{eq:t-Exact}),
is $t_{1}t_{3}/t_{2}$. This is also the effective hopping in the
SDRG scheme provided $\left|t_{2}\right|>\left|t_{1}\right|,\left|t_{3}\right|$.
On the other hand if, say, $\left|t_{1}\right|>\left|t_{2}\right|,\left|t_{3}\right|$,
then in the SDRG scheme used to derive Eq.\ (\ref{eq:P(eta)}), hoppings
$t_{1}$ and $t_{2}$ are decimated out and only $t_{3}$ remains
(as a consequence of open boundary conditions). Thus, the effective
hopping is $t_{3}$. If the hopping in the leads $t_{0}$ where included,
this problem would be avoided. However, other problems would appear,
such as, for instance, the definition of the chain length $L$.

In the logarithmic variable $\zeta$, the effective hopping is the
final position of a random walk after taking $L/2$ steps to the right
and $L/2-1$ to the left. In the SDRG scheme, the paths in which the
random walk cross the negative side are removed, as if there was a
hard wall at the origin. This is why the probability of finding the
random walker near the origin vanishes in the SDRG method, see Eq.\ (\ref{eq:P(eta)2}),
while it is maximum in the exact approach, see Eq.\ (\ref{eq:Q(eta)}).
For large $\eta$, the SDRG result agrees well with the exact one
if we identify $\sigma_{0}$ with $u_{0}$, as expected.

Despite the huge difference in the behavior of ${\cal P}\left(\eta\right)$
and ${\cal Q}\left(\eta\right)$ for $\eta\ll1$, the corresponding
beta functions in the Ohmic regime agree quite well with each other,
as shown in our analytical calculations and as can be seen in Fig.\ \ref{fig:The-beta-function}.
On the other hand, for the localized regime $\eta\gg1$, although
${\cal P}$ and ${\cal Q}$ agree remarkably well, surprisingly, the
corrections to the localized regime are quite different, as we have
shown analytically and can be clearly seen in Fig.\ \ref{fig:The-beta-function}.
The approach to the strongly localized regime $\beta=\frac{1}{2}\ln g$
is much faster in the SDRG method. We point out that this is not due
to the fact that $\eta$ can be negative in the exact calculation.
Recall that the transmittance in Eq.\ (\ref{eq:transmitancePH})
is an even function with respect to $\eta$. This remarkable agreement
only depends on the scaling of the variable $\zeta$ with the system
size $L$.

It is not our purpose here to modify the hierarchical decimation procedure
of the SDRG method in order to correctly handle the boundary conditions
analytically. Our main purpose is to show that the SDRG method can
be used to compute the beta function easily. Further developments
in higher dimensions will have to be tackled numerically since there
are very few analytical results using the SDRG scheme.\ \cite{motrunich-damle-huse-prb02,zhou-etal-epl09}
Besides, boundary conditions is higher dimensions are less important
and handling them can be easily accomplished via a numerical implementation
of the SDRG method.

\subsection{A simpler SDRG approach\label{sub:PH-simpleSDRG}}

We now introduce a different approach for computing the beta function
analytically in the framework of the SDRG method. As we discussed
before (Sec.\ \ref{sub:PH-SDRG}), it is not simple to compute the
distribution of the last hopping for a finite chain. Part of this
difficulty is due to the boundary conditions. As we expect this to
introduce little effect in the thermodynamic limit, we use a simpler
approach as explained below.

Consider an infinite chain. We run the SDRG method until the average
distance between the undecimated sites is $L$. At this stage, we
break the chain into pieces of consecutive sites, and consider each
piece as a representative of a finite chain of size $L$. Within this
simple approach, the distribution of the last hopping is exactly the
distribution of hoppings in Eq.\ (\ref{eq:PH-distribtion}), which
can be recast as 
\begin{equation}
{\cal P}\left(\eta\right)=e^{-\eta},\label{eq:PH-2}
\end{equation}
 with $\left(u_{0}+\Gamma\right)\eta=\ln\left(\Omega/t\right)$, and
the energy cutoff $\Gamma$ being related to $L$ via activated dynamical
scaling\ \cite{fisher95,hoyosvieiralaflorenciemiranda} 
\begin{equation}
L=\left(1+\Gamma/u_{0}\right)^{2}.\label{eq:L-Gamma-PH}
\end{equation}

It is clear that our boundary conditions are artificial and unlikely
correspond to a real physical situation. Moreover, Eq.\ (\ref{eq:PH-2})
is quite different from Eqs.\ (\ref{eq:P(eta)}) and (\ref{eq:P(eta)2}).
The largest difference is in the behavior for $\eta\rightarrow0$.
However, as we show below, this is of little importance for the \emph{average}
quantities.

We are now in a position to compute the beta function. The conductance
is $\ln g=\ln4-4\Gamma-2u_{0}-2\left\langle \ln\left(1-e^{-2\left(u_{0}+\Gamma\right)\eta-2\Gamma}\right)\right\rangle $
and $\beta=-\left(\Gamma+u_{0}\right)\left(2+\frac{{\rm d}}{{\rm d}\Gamma}\left\langle \ln\left(1-e^{-2\left(u_{0}+\Gamma\right)\eta-2\Gamma}\right)\right\rangle \right)$.
It is now clear that the beta function becomes universal only in the
limit of small disorder ($u_{0}\rightarrow0$) and large chains ($L\rightarrow\infty$)
but finite $u_{0}\sqrt{L}$. After performing the averages we obtain
\begin{eqnarray}
\ln g & = & \ln4-4\Gamma+2\sum_{n=1}^{\infty}e^{-2n\Gamma}/\left(n\left(2n\Gamma+1\right)\right),\label{eq:lng-PH-SDRG2}\\
\beta & = & -2\Gamma-4\Gamma\sum_{n=1}^{\infty}\left(n\Gamma+1\right)e^{-2n\Gamma}/\left(2n\Gamma+1\right)^{2}\!.\quad\label{eq:beta-PH-SDRG2}
\end{eqnarray}
Here, $\Gamma\rightarrow u_{0}\sqrt{L}$ becomes the parametrization
constant and depends on the combination $u_{0}\sqrt{L}$ as before.
For comparison, the corresponding beta function is plotted in Fig.\ \ref{fig:The-beta-function}
(dotted blue line). As expected, the agreement with the first analytical
SDRG approach is remarkable. Noticeable differences in $\beta$ arise
only when $g$ is of order unity.

The Ohmic regime is obtained in the limit $\Gamma\ll1$. Here, we
approximate the sums by integrals and find that $\beta\approx-1-1.856/\sqrt{g}$.
For the localized regime $\Gamma\gg1$, on the other hand, we keep
only the $n=1$ contribution in the sums. It follow that $\beta\approx\frac{1}{2}\ln(g/4)-\frac{1}{2}\sqrt{g}\left[1+4/\ln\left(g/4\right)\right]$.

The great advantage of this naive approach is its simplicity. It captures
the qualitative features of the beta function such as the localized
and Ohmic regime, and allows us to determine the conditions for universality.
This simplicity will come in handy when applying the method to the
generic case, as we do in the following.

\section{The generic case\label{sec:The-generic-case}}

In this section, we compute the beta function when particle-hole symmetry
is broken. As in Sec.\ \ref{sec:PH-symmetry}, we consider different
approaches and compare them.

\subsection{Numerically exact results}

Let us first consider the Hamiltonian in Eq.\ (\ref{eq:Hamiltonian})
with diagonal disorder only ($t_{i}=t_{0}=\Omega_{0}$). We will consider
$\varepsilon$'s that are symmetrically distributed around the origin
according to 
\begin{equation}
R_{0}\left(\varepsilon\right)=\Theta(\varepsilon_{0}-\left|\varepsilon\right|)\left(\varepsilon_{0}/\left|\varepsilon\right|\right)^{1-1/z}\left(\varepsilon_{0}z\right)^{-1},\label{eq:R0}
\end{equation}
where $\varepsilon_{0}$ is the maximum value of $\left|\varepsilon\right|$
and $z$ is an additional parameter.

We were not able to obtain exact analytical results for the beta function
in this case. Thus, we implemented numerically the transformations
in Eqs.\ (\ref{eq:t-tdec})---(\ref{eq:e-edec}) and computed the
beta function according to Eq.\ (\ref{eq:beta-function}). We considered
chains of length ranging from $L=10^{2}$ up to $10^{3}$, cutoff
energy from $\varepsilon_{0}=0.00625$ up to $0.8$, and four different
values of $z=10^{k},$ with $k=-2,\dots,1$. The different data sets
are shown in Fig.\ \ref{fig:The-beta-function}. For our discussion,
there is no need to distinguish the parameters used for each data
set. As can be seen, the beta function seems to be universal even
though the single-parameter scaling theory is not applicable to this
case. This is because the violations are quite small.\ \cite{schomerus-titov-prb03}

\begin{figure}[t]
\begin{centering}
\includegraphics[clip,width=1\columnwidth]{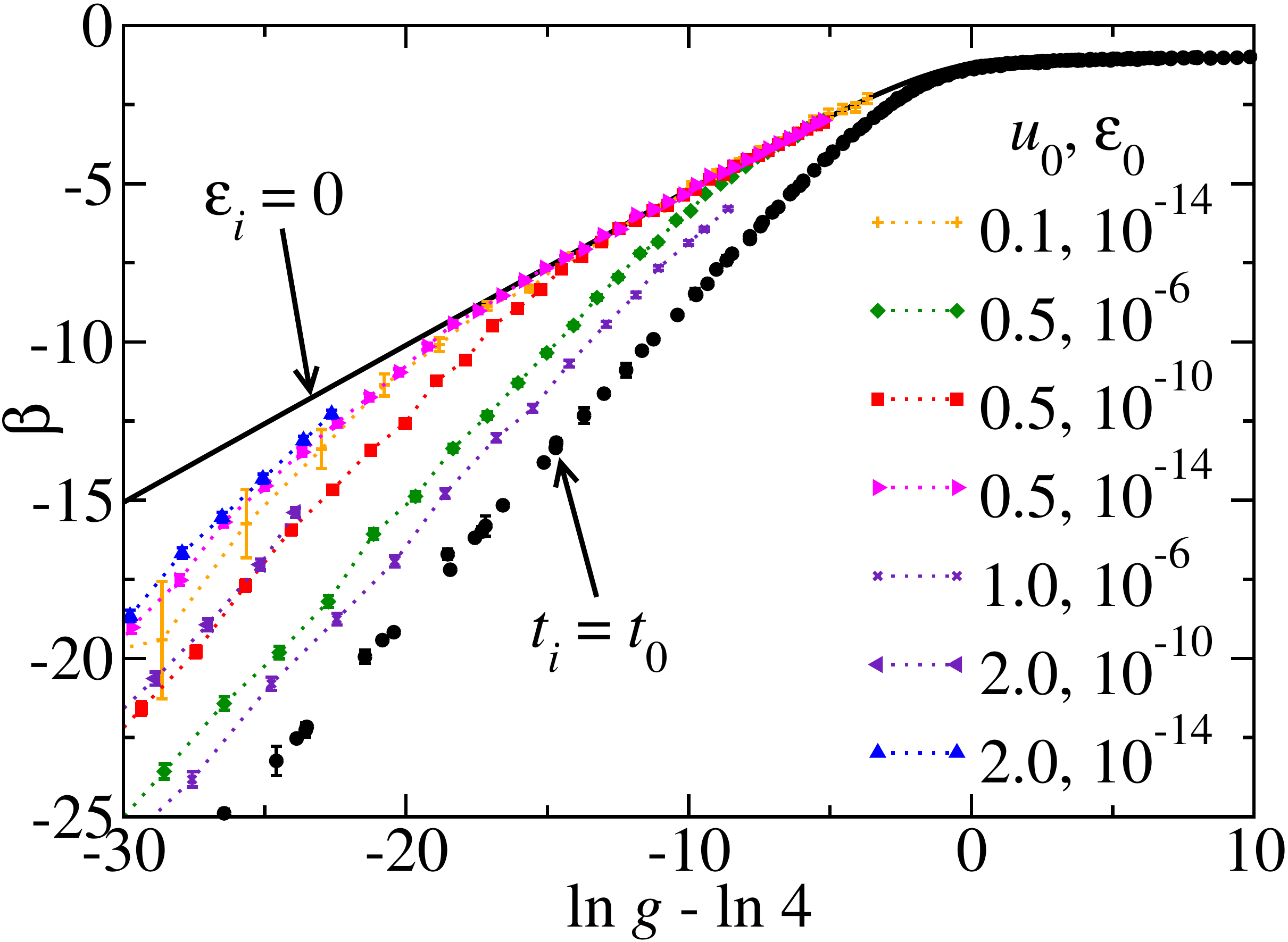}
\par\end{centering}

\protect\caption{The beta function for different disorder strengths, highlighting the
its non-universality when particle-hole symmetry ($\varepsilon_{i}=0$)
is broken.\label{fig:The-beta-function-2}}
\end{figure}

The beta function is clearly different from the particle-hole symmetric
case, however. The question we now address is how the particle-hole
symmetric behavior $\beta=\frac{1}{2}\ln g$ changes when this symmetry
is weakly broken by the introduction of small random $\varepsilon$'s.
Here, we study systems in which the $t$'s are distributed between
$0$ and $\Omega_{0}$ according to 
\begin{equation}
P_{0}\left(t\right)=\left(\Omega_{0}/t\right)^{1-1/u_{0}}\left(\Omega_{0}u_{0}\right)^{-1},\label{eq:P0}
\end{equation}
where $u_{0}$ parameterizes the disorder strength. For this case,
we considered chains of sizes varying from $L=10^{2}$ up to $10^{4}$,
onsite cutoff energies varying from $\varepsilon_{0}=10^{-14}$ to
$10^{-1}$, $\Omega_{0}=1$, disorder strengths $z=0.1,$ $1$, and
$10$, and $u_{0}=0.1$, $0.5$, $1$, and $2$. All chains have qualitatively
the same behavior. For clarity, we show only a few representative
ones in Fig.\ \ref{fig:The-beta-function-2} for $z=1$ (see also
App.\ \ref{sec:Alternative-conductance}). The continuous line is
the exact result for the particle-hole symmetric case, Eqs.\ (\ref{eq:lng-Exact})
and (\ref{eq:beta-Exact}). The black circles are for the limiting
case of uniform hopping discussed above (same data points of Fig.\ \ref{fig:The-beta-function}).
We discuss the observed non-universality of the beta function in the
following subsections.

\subsection{SDRG results\label{sub:Generic-SDRG}}

We now apply the analytical SDRG method to the general Hamiltonian
in Eq.\ (\ref{eq:Hamiltonian}) with both diagonal and off-diagonal
disorder. As expected, the RG flow equations are much harder to solve
due to the structure of Eqs.\ (\ref{eq:t-tdec})---(\ref{eq:e-edec}).
We can simplify them, however, by assuming that, near the fixed point,
the system disorder is so strong ($\Omega\gg|t_{i}|,|\varepsilon_{i}|$)
that those equations can be approximated by $\tilde{t}=t_{i}t_{j}/\Omega$
and $\tilde{\varepsilon}_{i}=\varepsilon_{i}$. In this approximation,
the signs of $t$'s and $\varepsilon$'s become irrelevant and we
will henceforth ignore them. With this, the flow equations for the
distributions $P\left(t\right)$ and $R\left(\varepsilon\right)$
become 

\begin{eqnarray}
\frac{\partial P}{\partial\Omega} & = & R(\Omega)P(t)-\left[P(\Omega)+R(\Omega)\right]P\stackrel{t}{\otimes}P\label{eq:flow-P}\\
\frac{\partial R}{\partial\Omega} & = & -R(\Omega)R(\varepsilon).\label{eq:flow-R}
\end{eqnarray}
where $P\stackrel{t}{\otimes}P=\int{\rm d}t_{1}{\rm d}t_{1}P(t_{1})P(t_{2})\delta\left(t-t_{1}t_{2}/\Omega\right)$.
The first terms on the right-hand sides come from the normalization
of $P$ and $R$ as $\Omega$ is lowered. The second term on the right-hand
side of Eq.\ (\ref{eq:flow-P}) implements Eqs.\ (\ref{eq:t-tdec})
and (\ref{eq:t-edec}). Notice that, at this very simple level of
approximation, the only renormalization on $R$ is due to the lowering
of the cutoff. Thus, any solution of the type $R\left(\varepsilon\right)=f\left(\varepsilon\right)A\left(\Omega\right)$,
where $f\left(\varepsilon\right)\geq0$ is any non-pathological distribution
function and $A\left(\Omega\right)=1/\int_{0}^{\Omega}{\rm d}\varepsilon f\left(\varepsilon\right)$
a normalization constant, is a solution to Eq.\ (\ref{eq:flow-R}).

Before presenting the fixed-point solution for $P\left(t\right)$,
notice the RG flow is quite similar to that of the random transverse-field
Ising chain deep in the paramagnetic Griffiths phase, if we associate
transverse fields with onsite energies and exchange couplings with
hoppings. In that case, asymptotically, only transverse fields are
decimated, thus renormalizing the coupling constants.\ \cite{fisher95}
That is exactly the asymptotic flow in Eqs.\ (\ref{eq:flow-P}) and
(\ref{eq:flow-R}). Even if initially most $t$'s are greater than
most $\varepsilon$'s, the initial flow of $P$ is towards the singular
distribution of Eq.\ (\ref{eq:PH-distribtion}). At some point, $R\left(\Omega\right)$
will become of the order of $P\left(\Omega\right)$. After this point,
the singularity of $P$ will be enhanced because the $\varepsilon$'s
will dominate over the $t$'s, just like in the SDRG flow of the paramagnetic
Griffiths phase.\ \cite{fisher95}

With this similarity in mind and considering the initial distributions
in Eqs.\ (\ref{eq:R0}) and (\ref{eq:P0}), we use the following
Ansatz for the unknown distributions

\begin{equation}
P(t)=\frac{1}{u_{\Omega}\Omega}\left(\frac{\Omega}{t}\right)^{1-\frac{1}{u_{\Omega}}},\quad R(\varepsilon)=\frac{1}{z\Omega^{\prime}}\left(\frac{\Omega^{\prime}}{\left|\varepsilon\right|}\right)^{1-\frac{1}{z}},\label{eq:fixed-points-P-R}
\end{equation}
where the hoppings and onsite energies are distributed between $0\leq t\leq\Omega$
and $0\leq\left|\varepsilon\right|\leq\Omega^{\prime}$, respectively.
We find that $\Omega^{\prime}=\min\left\{ \varepsilon_{0},\Omega\right\} $,
$u_{\Omega}=u_{0}+\Gamma$ for $\Omega>\varepsilon_{0}$ and $u_{\Omega}=\left(u_{0}+\Gamma_{\varepsilon}+z\right)e^{\left(\Gamma-\Gamma_{\varepsilon}\right)/z}-z,$
for $\Omega<\varepsilon_{0}$, where $\Gamma=\ln\left(\Omega_{0}/\Omega\right)$,
and $\Gamma_{\varepsilon}=\ln\left(\Omega_{0}/\varepsilon_{0}\right)$.
In fact, there are other fixed point solutions parameterized by different
$R\left(\varepsilon\right)$. However, we find that the important
features of the beta function are fairly independent of the family
of solutions chosen, as long as they are constrained by the condition
that $R\left(\Omega\right)\gg P\left(\Omega\right)$, for $\Omega$
less than the crossover energy scale $\Omega_{0}e^{-\Gamma^{*}}$,
where $\Gamma^{*}$ is defined later in connection with Eq.\ (\ref{eq:L-Gamma-Generic}).
Moreover, the solutions in Eq.\ (\ref{eq:fixed-points-P-R}) are
convenient because they recover Eq.\ (\ref{eq:PH-distribtion}) in
the limit $z\rightarrow\infty$ in a simple manner. We stress that
this freedom in choosing the function $R\left(\varepsilon\right)$
is a consequence of the approximations made in arriving at Eqs.~(\ref{eq:flow-P})
and (\ref{eq:flow-R}). The full SDRG flow, of course, determines
this function uniquely. Finally note that the flow described by Eq.~(\ref{eq:fixed-points-P-R})
clearly corresponds to a two-parameter scaling situation, $u_{0}$
and $z$ here playing the role of the two parameters.

As usual, we will focus on long chains that are weakly disordered.
This means that the $t$'s are narrowly distributed near $\Omega_{0}$
and the $\left|\varepsilon\right|$'s are much smaller than $\Omega_{0}$.
In this limit, the transmittance in Eq.\ (\ref{eq:transmitance})
reduces to the one in Eq.\ (\ref{eq:transmitancePH}). This means
that all the effects on the beta function due to the particle-hole
symmetry-breaking $\varepsilon$'s are encoded in the behavior of
the renormalized hopping $t_{1,L}$. In this case, the Ohmic regime
is straightforwardly recovered since it happens in the limit of $1\approx t_{1,L}\gg\left|\varepsilon_{1,2}\right|$.
The RG flow is just like the particle-hole symmetric case and Eq.\ (\ref{eq:fixed-points-P-R})
reduces to Eq.\ (\ref{eq:PH-distribtion}). We thus focus on the
localized regime where $t_{1,L}\ll\left|\varepsilon_{1,2}\right|\ll1$,
and $T\approx\left(2t_{1,L}\right)^{2}$. 

For that we need the distribution of $t_{1,L}$. As mentioned before,
the RG flow is like that of the paramagnetic Griffiths phase of the
random-transverse field Ising model and this distribution is known.\ \cite{fisher-young-RTFIM}.
Indeed, the family of solutions in Eq.\ (\ref{eq:fixed-points-P-R})
is analogous to the line of fixed-point distributions of the Griffiths
phase.\ \cite{fisher95} Thus, using the results of reference\ \onlinecite{fisher-young-RTFIM},
we have, in the limit $L\gg L^{*}$ and $L^{-z}\ll\varepsilon_{0}$,
\begin{equation}
{\cal P}\left(\zeta\right)=e^{-\left(\zeta-L/L^{*}\right)^{2}/\left(4L\right)}/\sqrt{4\pi L},\label{eq:P(zeta)-Generic-SDRG}
\end{equation}
where $\zeta=\ln(\Omega_{0}/t_{1,L})$ and $L^{*}\approx\left(\max\left\{ \Gamma_{\varepsilon},z\right\} /u_{0}\right)^{2}$
is a crossover length which will be discussed later in connection
with Eq.\ (\ref{eq:L-Gamma-Generic}). This can be understood as
follows. For $L\gg L^{*}$, the later stages of the RG flow are dominated
by $\varepsilon$'s decimations {[}see Eqs.\ (\ref{eq:t-edec}) and
(\ref{eq:e-edec}){]}. Thus, $\tilde{\zeta}$ renormalizes in a simple
additive fashion and we expect the mean value of $\zeta$ to be proportional
to $L$ and its variance to follow the central limit theorem. We note
that the derivation of Eq.\ (\ref{eq:P(zeta)-Generic-SDRG}) is fairly
non-trivial\ \cite{fisher-young-RTFIM} and corrections due to a
finite $u_{0}$ are quite involved. We will come back to these features
later when we discuss the simpler SDRG approach. Since $\ln g_{s}\approx-2\zeta$,
the $g_{s}$ distribution in this limit is log-normal, with the average
and variance of $\ln g_{s}$ scaling linearly with $L$.

We finally note that, although ${\cal P}\left(\zeta\right)$ depends
on a non-universal constant $L^{*}$, the localized regime is universal:
$\ln g\approx\ln4-2\left\langle \zeta\right\rangle =\ln4-2L/L^{*}$,
and $\beta\approx-2L/L^{*}\approx\ln g$. This is the more familiar
result $\beta\sim\ln g$ for localized states in the strongly localized
regime. In this limit, $\beta$ is universal, as seen in the data
of Fig.\ \ref{fig:The-beta-function}. Non-universal corrections
will be dealt with in the following.

\subsection{A simpler SDRG approach\label{sub:Generic-simpleSDRG}}

We now generalize the simpler approach of Section~\ref{sub:PH-simpleSDRG}
to the generic case without particle-hole symmetry. We thus take the
distribution of hoppings $t_{1,L}$ and onsite energies $\varepsilon_{1}$
and $\varepsilon_{2}$ in the final link to be given by the bulk distributions
of Eq.\ (\ref{eq:fixed-points-P-R}), which can be recast as 
\begin{equation}
{\cal P}\left(\eta\right)=\Theta\left(\eta\right)e^{-\eta},\mbox{ and }{\cal R}\left(\chi\right)=\Theta\left(\chi-\chi_{0}\right)z^{-1}e^{-\chi/z},\label{eq:P-R-Generic-simpleSDRG}
\end{equation}
where $\eta=\ln\left(\Omega/t\right)/u_{\Gamma}$, $\chi=\ln\left(\Omega/\left|\varepsilon\right|\right)$,
$\chi_{0}=\ln\left(\Omega/\Omega^{\prime}\right)$. As discussed in
Section~\ref{sub:Generic-SDRG}, we can use in this case the transmittance
given in Eq.\ (\ref{eq:transmitancePH}). Parameterizing the conductance
and the beta function through $\Gamma=\ln\left(\Omega_{0}/\Omega\right)$
\begin{eqnarray}
\ln g & = & \ln4-2\Gamma-2u_{\Gamma}-2\left\langle \ln\left(1-e^{-2\left(\eta u_{\Gamma}+\Gamma\right)}\right)\right\rangle ,\\
\beta & = & -2\left(1+\dot{u}_{\Gamma}+2\left\langle \frac{\eta\dot{u}_{\Gamma}+1}{e^{2\left(\eta u_{\Gamma}+\Gamma\right)}-1}\right\rangle \right)\frac{{\rm d}\Gamma}{{\rm d}\ln L},
\end{eqnarray}
where $\dot{u}_{\Gamma}=\frac{{\rm d}u_{\Gamma}}{{\rm d}\Gamma}=1$
and $\frac{{\rm d}\Gamma}{{\rm d}\ln L}=u_{\Gamma}/2$ for $\Gamma<\Gamma_{\varepsilon}$,
and $\dot{u}_{\Gamma}=1+u_{\Gamma}/z$ and $\frac{{\rm d}\Gamma}{{\rm d}\ln L}=\frac{u_{\Gamma}z}{u_{\Gamma}+2z}$
for $\Gamma>\Gamma_{\varepsilon}$. Note that, in this simpler approach,
we obtain the relation between the energy cutoff $\Gamma$ and the
chain length $L$ by taking the latter to be the mean distance between
active sites. From the rate equation ${\rm d}n=n\left(2P\left(\Omega\right)+R\left(\Omega\right)\right){\rm d}\Omega$,
where $n=L^{-1}$ is the density of active sites in the effective
chain, we find 
\begin{equation}
L=\left(u_{\Gamma}/u_{0}\right)^{2}\min\left\{ 1,e^{-\left(\Gamma-\Gamma_{\varepsilon}\right)/z}\right\} .\label{eq:L-Gamma-Generic}
\end{equation}
In the limit $z\rightarrow\infty$ (or for $\Gamma<\Gamma_{\varepsilon}$),
the activated dynamical scaling of Eq.\ (\ref{eq:L-Gamma-PH}) is
recovered. For $\Gamma-\Gamma_{\varepsilon}\gg z$, on the other hand,
the usual power-law dynamical scaling $\Omega\sim L^{-z}$ holds,
with $z$ playing the role of the dynamical exponent. The crossover
between the two regimes happens when the RG flow is dominated by both
$t$- and $\varepsilon$-decimations, i.e., when the typical values
of $t$'s and $\varepsilon$'s are of same order, namely, when $u_{\Gamma^{*}}=z$.
Thus, the crossover energy scale is $\Gamma^{*}=\Gamma_{\varepsilon}+\max\left\{ 0,z\ln\left(2z/\left(u_{0}+\Gamma_{\varepsilon}+z\right)\right)\right\} $,
which gives a crossover length scale $L^{*}\approx\left(\max\left\{ z,\Gamma_{\varepsilon}\right\} /u_{0}\right)^{2}$.

Averaging over $\eta$, we get 
\begin{eqnarray}
\ln g & = & \ln4-2\Gamma-2u_{\Gamma}+2\sum_{n=1}^{\infty}\frac{e^{-2n\Gamma}}{nf_{n,\Gamma}},\label{eq:lng-SDRG-generic}\\
\beta & = & -2\left(1+\dot{u}_{\Gamma}+2\sum_{n=1}^{\infty}\frac{g_{n,\Gamma}e^{-2n\Gamma}}{f_{n,\Gamma}^{2}}\right)\frac{{\rm d}\Gamma}{{\rm d}\ln L},\label{eq:beta-SDRG-generic}
\end{eqnarray}
where $f_{n,\Gamma}=2nu_{\Gamma}+1$ and $g_{n,\Gamma}=\left(f_{n,\Gamma}+\dot{u}_{\Gamma}\right)/f_{n,\Gamma}^{2}$.

We now discuss some limits of interest. Since $\epsilon_{0}\ll\Omega_{0}$,
our calculation is valid only in the limit $\Gamma_{\varepsilon}\gg1$.
In the particle-hole symmetric case, universality comes about in the
limit $L\rightarrow\infty$ and $u_{0}\rightarrow0$ with $\Gamma=u_{0}\sqrt{L}$
finite. In order to recover the Ohmic regime ($\Gamma\ll1$), then
$u_{\Gamma}=\Gamma+{\cal O}\left(\Gamma^{2}/z\right)$ and Eqs.\ (\ref{eq:lng-SDRG-generic})
and (\ref{eq:beta-SDRG-generic}) become $z$-independent recovering
Eqs.\ (\ref{eq:lng-PH-SDRG2}) and (\ref{eq:beta-PH-SDRG2}). We
then conclude that the Ohmic regime is the same in both particle-hole
symmetric and generic cases.

The limit $z\rightarrow\infty$ is straightforward. It recovers the
particle-hole symmetric case $\beta=\frac{1}{2}\ln g$ simply because
$R\left(\varepsilon\right)$ becomes extremely singular and the $\varepsilon$'s
essentially play no role in the RG flow.

Let us now focus on the localized regime for finite $z$. When $1\ll\Gamma\ll\Gamma^{*}$,
we are back to the particle-hole symmetric case where $\beta\approx\frac{1}{2}\ln g$.
When $\Gamma$ becomes greater than $\Gamma^{*}$, then the RG flow
veers from the particle-hole symmetric one and we expect $\beta$
to cross over to the $\sim\ln g$ behavior. In the limit $\Gamma-\Gamma^{*}\gg z$,
we find that 
\begin{equation}
\beta=\ln\left(g/4\right)+2\Gamma^{*}+16z^{2}/\ln\left(g/4\right),\label{eq:beta-SDRG-simple-Generic}
\end{equation}
up to corrections of order ${\cal O}\left[z\ln\left(\ln\left(4/g\right)\right)\right].$
Notice that $\beta$ is non-universal due to the constant $\Gamma^{*}$.
As we have mentioned before, this result can be interpreted as a sharp
crossover from the universal particle-hole symmetric case $\frac{1}{2}\ln\left(g/4\right)$
to the generic one $\ln(g/4)+{\rm const}$. For $\beta$ to be continuous,
the constant must be $-\frac{1}{2}\ln\left[g\left(\Gamma^{*}\right)/4\right]$
which, according to Eq.\ (\ref{eq:lng-PH-SDRG2}), is $\sim2\Gamma^{*}$.

This result explains the non-universal beta function found numerically
in Fig.\ \ref{fig:The-beta-function-2}. We note that the crossover
constant $\Gamma^{*}$ does not fit quite well the numerical data
in Fig.\ \ref{fig:The-beta-function-2} for a very simple reason.
In our analytical approach, we have neglected the corrections to the
renormalization of $\varepsilon$'s in Eq.\ (\ref{eq:e-tdec}). These
corrections enhance the bare cutoff $\varepsilon_{0}$ and, consequently,
the crossover energy $\Gamma^{*}$ will be slightly suppressed, in
agreement with the exact numerical result.

Finally, we comment on the beta function for the uniform hopping case.
Unfortunately, our analytical approach cannot be used in that case
because we neglected the $\varepsilon$-corrections in Eq.\ (\ref{eq:t-tdec}).
As a result, all the hoppings retain the value $t_{0}$ along the
RG flow, which then fixes the cutoff energy at $\Omega=\Omega_{0}$
and the localized regime is never reached.

\section{The Lyapunov exponent\label{sec:The-Lyapunov-exponent}}

In this Section, we briefly comment on the Lyapunov exponent. It is
usually defined as $\gamma=\lim_{L\rightarrow\infty}\gamma_{L}$,
where 
\begin{equation}
\gamma_{L}=\frac{\ln\left(1+g^{-1}\right)}{L}=-\frac{\left\langle \ln\left(T\right)\right\rangle }{L},\label{eq:Lyapunov-usual}
\end{equation}
where $T$ is the transmittance {[}see Eq.\ (\ref{eq:transmitance}){]}.
Single-parameter scaling theory\ \cite{anderson-etal-prb80} predicts
that the standard deviation of $-\frac{\ln T}{L}$, which we will
call $\sigma_{L}$ is such that 
\begin{equation}
\sigma_{L}^{2}\approx2L^{-2}\ln\left(\cosh\gamma L\right)\rightarrow2\gamma/L.\label{eq:sigmaL-usual}
\end{equation}
The fact that $\sigma_{L}^{2}$ vanishes as $L^{-1}$ is a consequence
of the central limit theorem, which follows from the hypothesis of
phase randomization of the single-parameter scaling theory.

For states near the band center, it is known that single-parameter
scaling theory is not valid.\ \cite{schomerus-titov-prb03,deych-etal-prl03}
Actually, a two-parameter scaling theory is needed.\ \cite{balents-fisher-prb97}.
However, for the particle-hole symmetric case, another single-parameter
scaling theory is possible. Evidently, the definition in Eq.\ (\ref{eq:Lyapunov-usual})
cannot be used in this case. As shown in Eqs.\ (\ref{eq:P(eta)}),
(\ref{eq:P(eta)2}), (\ref{eq:Q(eta)}) and (\ref{eq:PH-2}), the
correct scaling variable is $\left\langle \ln T\right\rangle /\sqrt{L}$.
Thus, the useful definition for the Lyapunov exponent should be 
\begin{equation}
\sqrt{\gamma_{L}}=\frac{\left\langle \ln T\right\rangle }{\sqrt{L}},\label{eq:Lyapunov-PH}
\end{equation}
which can be easily computed. Using the exact result in Eq.\ (\ref{eq:Q(eta)}),
$\left\langle \ln T\right\rangle =\ln4-2\left\langle \ln\left(1+e^{-2\alpha\eta}\right)\right\rangle ,$
with $\alpha=\sigma_{0}\sqrt{L/2}$. For $\alpha\rightarrow\infty,$
the integral is dominated by the negative values of $\eta$. Then,
$\ln\left(1+e^{-2\alpha\eta}\right)\approx-2\alpha\eta$ (for $\eta<0$),
yielding $\left\langle \ln T\right\rangle =\ln4-4\alpha/\sqrt{\pi}+{\cal O}\left(\alpha^{-2}\right)$.
Therefore, $\gamma_{L}=8\sigma_{0}^{2}/\pi$. Recall that the wave
function is stretched-exponentially localized,\ \cite{inui-trugman-abrahams-prb94}
$\ln\left|\psi\right|^{2}\sim-\sqrt{\gamma L}$, and the localization
length is the inverse of the Lyapunov exponent 
\begin{equation}
\gamma^{-1}=\pi/(8\sigma_{0}^{2}).\label{eq:Lloc-PH}
\end{equation}

In the same manner, the variance of $\ln T$ in the $\alpha\rightarrow\infty$
limit is $\sigma_{\ln T}^{2}=16\alpha^{2}\left(1-\pi^{-1}\right)$.
We then conclude that the single-parameter scaling theory for the
particle-hole symmetric dictates that 
\begin{equation}
\sigma_{L}^{2}=\frac{16\alpha^{2}}{L}\left(1-\pi^{-1}\right)=\left(\pi-1\right)\gamma.\label{eq:sigmaL-PH}
\end{equation}
The remarkable difference from the generic case Eq.\ (\ref{eq:sigmaL-usual})
is that $\sigma_{L}$ does not vanish in the thermodynamic limit,
i.e., the quantity $\ln T/\sqrt{L}$ is not self-averaging.\ \cite{soukoulis-economou-prb81}
This is the hallmark of the infinite-randomness fixed point physics
of the particle-hole symmetric case.\ \cite{fisher95}

For completeness, let us analyze the generic case using the SDRG approach
of Sec.\ \ref{sub:Generic-SDRG}. It is clear from the distribution
in Eq.\ (\ref{eq:P(zeta)-Generic-SDRG}) that $\gamma=1/L^{*}$ and
that $\sigma_{L}^{2}=2/L=2L^{*}\gamma/L$. In contrast to Eq.\ (\ref{eq:sigmaL-usual}),
the ratio $L\sigma_{L}^{2}/\gamma$ is not universal, as expected
from the two-parameter scaling theory. Indeed, even for the special
case of uniform hopping ($t_{i}=t_{0}$), the beta function is not
universal, even though the non-universalities are hard to characterize
numerically, since single-parameter scaling is only weakly violated.\ \cite{schomerus-titov-prb03}
As shown in Ref. \onlinecite{schomerus-titov-prb03}, the ratio $L\sigma_{L}^{2}/\gamma$
when $L\to\infty$ is actually $2.094$, not $2$ as in Eq.\ (\ref{eq:sigmaL-usual}).

Finally, we note that the results of the simpler SDRG approach in
Sec.\ \ref{sub:Generic-simpleSDRG} are not accurate for computing
$\sigma_{L}^{2}$. Although the distribution in Eq.\ (\ref{eq:P-R-Generic-simpleSDRG})
has the correct scaling for the average $\left\langle \ln t_{1,L}\right\rangle $
(and thus, the correct scaling for the beta function), it overestimates
the fluctuations of $\ln t_{1,L}$, yielding the unphysical result
$\sigma_{\ln T}^{2}\sim L^{2}$.

\section{Conclusions and discussion\label{sec:Conclusion}}

In this paper, we have confirmed that a single-parameter scaling theory
is applicable to the particle-hole symmetric state of the one-dimensional
tight-binding model, yielding a universal beta function which, in
the localized regime, is $\beta=\frac{1}{2}\ln\left(g/4\right)$.
When particle-hole symmetry is broken by weak onsite disorder, the
band-center state displays a non-universal beta function, which crosses
over to $\beta=\ln g+{\rm const}$, with a non-universal constant.
As explained in the Introduction, even in this two-parameter scaling
case the leading term of $\beta$ is still universal and does not
depend on the particular definition of the conductance $g$ {[}either
Eq.\ (\ref{eq:g-average}) or Eq.\ (\ref{eq:g-alternative}){]}.
On the other hand, the non-universal sub-leading term, which comes
from corrections to scaling, does depend on the definition of $g$. 

All these conclusions can be interpreted in a simple way. Within the
SDRG method, the RG flow for the particle-hole symmetric case is identical
to that of the critical point of the random-transverse field Ising
chain, which is governed by an infinite-randomness fixed point which
is reflected in the novel single-parameter scaling theory. By introducing
onsite disorder, the SDRG flow deviates from the critical one towards
a line of fixed point that holds close analogy to the line of fixed
points of the Griffiths phase of the aforementioned Ising model. This
is an alternative interpretation of the two-parameter scaling theory
of the tight-binding model close to the band center.

One question that arises from these conclusions is why single-parameter
scaling holds for the particle-hole symmetric case while it is violated
for the generic case. In order to shed some light into this question,
we investigate whether the criterion for single-parameter scaling
developed in Refs.\ \onlinecite{deych-lisyansky-altshuler-prl00,deych-lisyansky-altshuler-prb01}
is violated in the particle-hole symmetric case. As we show shortly
below, the criterion is indeed violated and single-parameter scaling
should be violated. Of course, one has to be very careful in blindly
applying this criterion for the particle-hole symmetric case. As pointed
out by the authors, this additional symmetry introduces further complications
and their result cannot be directly applied here. In any case, a new
criterion for single-parameter scaling is desirable.

In Refs.\ \onlinecite{deych-lisyansky-altshuler-prl00,deych-lisyansky-altshuler-prb01},
it was stated that single-parameter scaling is valid as long as the
localization length is greater than $l_{s}=1/\sin\left(\pi N\left(E\right)\right)$,
where $N\left(E\right)$ is integrated the density of states at energy
$E$ normalized by the total number of states in that band. This should
be viewed as a necessary condition ensuring that the localization
length is greater than all other length scales in the system. In this
case, the phase randomization hypothesis of Ref.\ \onlinecite{anderson-etal-prb80}
would hold. For band edge states, $N\left(E\right)\ll1$ and thus
$l_{s}$ diverges. That is the reason why band edge states violate
the single-parameter scaling. For middle-of-the-band states, $N\left(E\right)\sim1/2$
and $l_{s}$ is microscopic. Thus, single-parameter scaling holds.
As pointed out in Ref.\ \onlinecite{deych-etal-prl03}, for the Hamiltonian
in Eq.\ (\ref{eq:Hamiltonian}), there are actually two bands, and
the $E=0$ state is actually a band-edge state between the two bands.
It was then shown that $l_{s}$ is indeed greater than the localization
length.

The integrated density of states for the particle hole symmetric case
was computed in Ref.\ \onlinecite{eggarter-riedinger-prb78}. Keeping
in mind that actually there are two bands, for $\left|E\right|\ll t_{0}$
and in the limit of small disorder $\sigma_{0}$, it is found that
$N\left(E\right)\approx1-\sigma_{0}^{2}[2\ln(t_{0}/E)]^{-2}$. Therefore,
the ratio between the localization length and $l_{s}$ is $\sim\pi^{2}/(32[\ln(t_{0}/E)]^{-2})$,
which vanishes logarithmically in the $E\rightarrow0$ limit. Thus,
the criterion of Refs.\ \onlinecite{deych-lisyansky-altshuler-prl00,deych-lisyansky-altshuler-prb01}
is also violated and single-parameter scaling is not possible. 

We finally conclude by recalling that the SDRG method here presented
can be applied to higher dimensions (see App.\ \ref{sec:SDRG}).
Although an analytical solution seems to be impossible, a numerical
implementation is possible and convenient due to the low numerical
cost of the method. This study will be undertaken in a future publication.
\begin{acknowledgments}
This work was supported by the NSF under Grants DMR-1005751 and PHYS-1066293,
by the Simons Foundation, by FAPESP under Grants 07/57630-5 and 2013/09850-7,
and by CNPq under Grants 304311/2010-3, 590093/2011-8 and 305261/2012-6.
We acknowledge the hospitality of the Aspen Center for Physics.
\end{acknowledgments}

\appendix

\section*{Appendices:}

\section{The SDRG recursion relations\label{sec:SDRG}}

In this appendix, we derive explicitly the SDRG transformations.

\subsection{Decimating a hopping term}

Consider the case when $\left|t_{23}\right|$ is the largest energy
scale of the problem. Then we treat

\[
H_{0}=\varepsilon_{2}c_{2}^{\dagger}c_{2}+t_{23}(c_{2}^{\dagger}c_{3}+{\rm h.c.})+\varepsilon_{3}c_{3}^{\dagger}c_{3},
\]
exactly and 

\[
H_{1}=\sum_{i\neq2,3}t_{2,i}(c_{2}^{\dagger}c_{i}+{\rm h.c.})+t_{3,i}(c_{3}^{\dagger}c_{i}+{\rm h.c.}),
\]
in second-order perturbation theory. Here we have assumed, for generality,
that all possible hoppings to sites 2 and 3 can occur.

In the occupation number basis of sites 2 and 3, $\left|1,0\right\rangle $
and $\left|0,1\right\rangle $, the unperturbed Hamiltonian is $H_{0}=\left(\begin{array}{cc}
\varepsilon_{2} & t_{23}\\
t_{23} & \varepsilon_{3}
\end{array}\right)$ with eigenenergies $\lambda_{\pm}=\bar{\varepsilon}\pm\sqrt{t_{23}^{2}+(\frac{\Delta\varepsilon}{2})^{2}}$,
where $2\bar{\varepsilon}=\varepsilon_{2}+\varepsilon_{3}$ and $\Delta\varepsilon=\varepsilon_{2}-\varepsilon_{3}$.
The corresponding eigenvectors are $\left|+\right\rangle =\beta\left|1,0\right\rangle +\alpha\left|0,1\right\rangle $
and $\left|-\right\rangle =\alpha\left|1,0\right\rangle -\beta\left|0,1\right\rangle $,
with $\alpha=t_{2,3}/\sqrt{(\lambda_{+}-\varepsilon_{3})^{2}+t_{2,3}^{2}}$
and $\beta=\lambda_{+}-\varepsilon_{3}/\sqrt{(\lambda_{+}-\varepsilon_{3})^{2}+t_{2,3}^{2}}.$
These are the two states we want to integrate out because they are
very distant from the band center, i.e. , $\lambda_{\pm}\approx\pm t_{23}$
which is far from $E=0$. The state $\left|0\right\rangle \equiv\left|0,0\right\rangle $
(corresponding to the particle being elsewhere), is the one we want
to keep. We now treat $H_{1}$ perturbatively. Since there is no correction
to first order ($\left\langle 0\left|H_{1}\right|0\right\rangle =0$),
we must go to second order. We use the notation $\left|1_{k};0\right\rangle $
to denote that the particle is at site $k$ different from sites 2
and 3, and $\left|0;\pm\right\rangle $ to the denote the two high-energy
states in which the particle occupies sites 2 and 3.

The onsite energy corrections are

\begin{eqnarray}
\delta\varepsilon_{k} & = & \sum_{s=\pm}\frac{\left|\left\langle 1_{k};0\left|H_{1}\right|0;s\right\rangle \right|^{2}}{-\lambda_{s}},
\end{eqnarray}
and the effective onsite energy then becomes 
\begin{equation}
\tilde{\varepsilon}_{k}=\varepsilon_{k}-\frac{\varepsilon_{3}t_{2,k}^{2}-2t_{2,3}t_{2,k}t_{3,k}+\varepsilon_{3}t_{3,k}^{2}}{t_{2,3}^{2}-\varepsilon_{2}\varepsilon_{3}},\label{eq:etilde-t-transf}
\end{equation}
which reduces to Eq.\ (\ref{eq:e-tdec})\textcolor{black}{{} in the
special }case of one dimension and nearest-neighbor hopping only. 

The effective hopping between sites $k$ and $l$ is

\begin{eqnarray}
\delta t_{k,l} & = & \sum_{s=\pm}\frac{\left\langle 1_{k};0\left|H_{1}\right|0;s\right\rangle \left\langle 0;s\left|H_{1}\right|1_{l};0\right\rangle }{-\lambda_{s}},
\end{eqnarray}
which gives 
\begin{equation}
\tilde{t}_{k,l}=t_{k,l}+\frac{\varepsilon_{3}t_{2,k}t_{2,l}-t_{2,3}(t_{2,k}t_{3,l}+t_{2,l}t_{3,k})+\varepsilon_{2}t_{3,k}t_{3,l}}{t_{2,3}^{2}-\varepsilon_{2}\varepsilon_{3}},\label{eq:ttilde-t-transf}
\end{equation}
which reduces to Eq.\ (\ref{eq:t-tdec}) when there are only nearest-neighbor
hoppings. 

Finally, the effective Hamiltonian is that given by Eq.\ (\ref{eq:Hamiltonian})
with the renormalized couplings $\tilde{\varepsilon}_{k}$ and $\tilde{t}_{k,l}$.
Notice there is no global shift in the energy. This is important because
we are looking for a good approximation for the state at zero energy.
Therefore, shifts in the energy coming from perturbation theory, which
are commonly disregarded in SDRG treatments, cannot be ignored here.

\subsection{Decimating an onsite energy term}

Consider now the case in which the highest energy scale is given by
an onsite energy, say $\left|\varepsilon_{2}\right|.$ In this case,
the states of interest are $\left|0\right\rangle \equiv\left|0_{2}\right\rangle $
and $\left|1\right\rangle \equiv\left|1_{2}\right\rangle $, representing
no and one particle on site 2, respectively, which are eigenstates
of $H_{0}=\varepsilon_{2}c_{2}^{\dagger}c_{2}$. Now, we treat $H_{1}=\sum_{k}t_{2,k}(c_{2}^{\dagger}c_{k}+{\rm h.c.})$
perturbatively. We thus discard state $\left|1\right\rangle $ because
it is too far from the reference energy $E=0$. The discarded state
corresponds to a particle strongly localized at site 2.

To first order of perturbation theory, there is no correction: $\left\langle 0\left|H_{1}\right|0\right\rangle =0$.
To second order, the matrix elements are 

\[
\tilde{H}_{k,l}=\frac{\left\langle 0_{2};1_{k}\left|H_{1}\right|1_{2};0\right\rangle \left\langle 1_{2};0\left|H_{1}\right|0_{2};1_{l}\right\rangle }{-\varepsilon_{2}}.
\]
Thus, the effective onsite energy is 

\begin{equation}
\tilde{\varepsilon}_{k}=\varepsilon_{k}-\frac{t_{2,k}^{2}}{\varepsilon_{2}},\label{eq:etilde-e-transf}
\end{equation}
and the effective hopping is 

\begin{equation}
\tilde{t}_{k,l}=t_{k,l}-\frac{t_{k,2}t_{2,l}}{\varepsilon_{2}}.\label{eq:ttilde-e-transf}
\end{equation}

As in the hopping transformation, there is no global shift in the
energy. Equations (\ref{eq:etilde-e-transf}) and (\ref{eq:ttilde-e-transf})
reduce to Eqs.\ (\ref{eq:e-edec}) and (\ref{eq:t-edec}), respectively.

\section{Local transformations in the transfer matrix formalism\label{sec:Transfer-matrix}}

For a 1D system with nearest-neighbor hopping only, the conductance
can be obtained from the product of the transfer matrices $T_{L}T_{L-1}...T_{2}T_{1}$
where $T_{i}=\left(\begin{array}{cc}
\frac{E-\varepsilon_{i}}{t_{i}} & -\frac{t_{i-1}}{t_{i}}\\
1 & 0
\end{array}\right)$, and $E$ is the eigenenergy (see, for instance, Ref.\ \onlinecite{pendry-jpc82}).
This multiplicative structure can be treated within the SDRG philosophy.
In the case one wants to integrate out site 2, for instance, then
it is natural to replace the product $T_{3}T_{2}T_{1}$ by $\tilde{T}_{3}\tilde{T}_{1}$.
Setting $E=0$, it is easy to show that the effective onsite energies
$\tilde{\varepsilon}_{1,3}$ and hopping $\tilde{t}_{1,3}$ of matrices
$\tilde{T}_{1,3}$ are exactly given by Eqs.\ (\ref{eq:e-edec})
and (\ref{eq:t-edec}), respectively. This also holds for the $t$-transformation,
where we compare $T_{4}T_{3}T_{2}T_{1}$ with $\tilde{T}_{4}\tilde{T}_{1}$.
Again, $\tilde{t}_{1,4}$ and $\tilde{\varepsilon}_{1,4}$ are given
by Eqs.\ (\ref{eq:t-tdec}) and (\ref{eq:e-tdec}), respectively.
These are surprising results, since Eqs.\ (\ref{eq:t-tdec}) and
(\ref{eq:e-tdec}) were obtained in second-order perturbation theory,
whereas the transfer matrix result is exact. We thus conclude that,
for the purpose of computing transport properties such as the conductance
or the beta function, the SDRG transformations Eqs.\ (\ref{eq:t-tdec})---(\ref{eq:e-edec})
are exact transformations yielding exact results.

However, this not true of other properties. For example, we have checked
via exact diagonalization of small chains that the spectrum obtained
via the SDRG method is not exact. It is not clear why the SDRG method,
which is based on second-order perturbation theory, reproduces exactly
the transfer matrix result. We conjecture that this is due to current
conservation in one dimension. The SDRG transformations of Eqs.\ (\ref{eq:etilde-t-transf})---(\ref{eq:ttilde-e-transf})
can in principle be applied to any geometry in any dimension. There
is no reason to believe that they are exact transformations in higher
dimensions, even for conducting properties. Unlike in one dimension,
there are many paths the current can take from one point to another
without necessarily going through a certain site or bond that has
been integrated out. In fact, interference effects that occur when
the current passes through more than one site are not likely to be
exactly captured by the local perturbative SDRG approach. Nevertheless,
in the limit of very strong disorder, the SDRG is expected to become
highly accurate.

\begin{figure}[b]
\begin{centering}
\includegraphics[bb=0bp 24bp 713bp 534bp,clip,width=1\columnwidth]{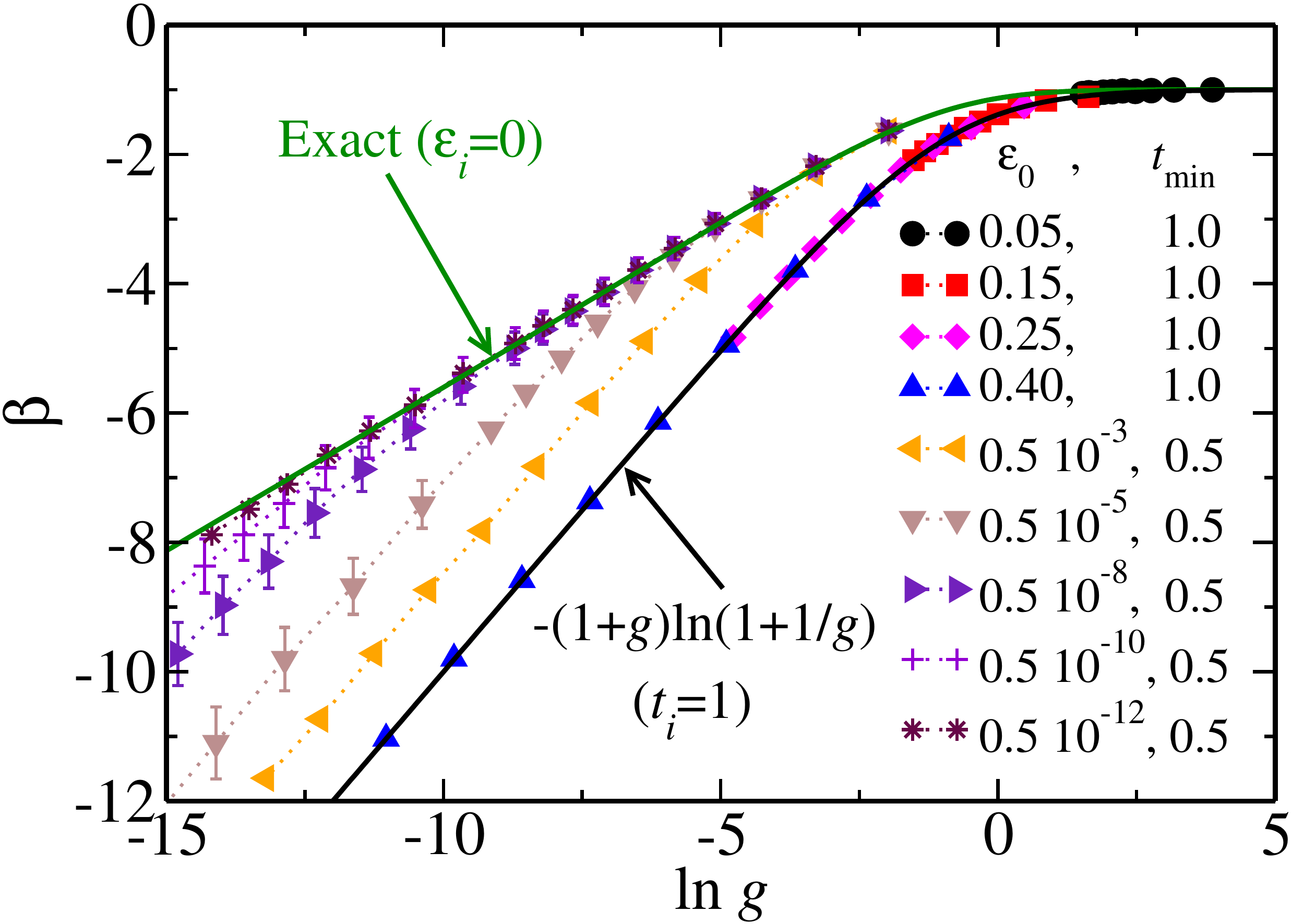}
\par\end{centering}

\protect\caption{The beta function calculated using $g=\left(T\right)_{{\rm geo}}/[1-\left(T\right)_{{\rm geo}}]$
for both diagonal and off-diagonal disorder. For comparison, we plot
the scaling form $-\left(1+g\right)\ln\left(1+g^{-1}\right)$ derived
in Ref.\ \onlinecite{anderson-etal-prb80}. Dotted lines are guides
to the eyes. \label{fig:beta-g-alternative}}
\end{figure}

\section{Alternative definitions of the conductance\label{sec:Alternative-conductance}}

We could have obtained the conductance through the definition proposed
in Ref.\ \onlinecite{anderson-etal-prb80}

\begin{equation}
g=e^{\left\langle \ln T\right\rangle }/\left(1-e^{\left\langle \ln T\right\rangle }\right).\label{eq:g-alternative}
\end{equation}
In Fig.\ \ref{fig:beta-g-alternative}, we plot the corresponding
beta function for system sizes varying from $L=10^{2}$ up to $10^{3}$,
for hoppings uniformly distributed between $t_{{\rm min}}<t<1.0$,
and onsite energies uniformly distributed between $-\varepsilon_{0}<\varepsilon<\varepsilon_{0}$.
 The exact curve for the particle-hole symmetric case ($\varepsilon_{i}=0$)
is obtained by doing the average in $\left\langle \ln T\right\rangle $
with the distribution in Eq.\ (\ref{eq:Q(eta)}). For the uniform
hopping case $t_{\mathrm{min}}=1.0$ (i.e., $t_{i}=1.0$), the beta
function $\beta=-\left(1+g\right)\ln\left(1+g^{-1}\right)$ of Ref.\ \onlinecite{anderson-etal-prb80}
seems to be recovered. However, this agreement is not perfect, since
single-parameter scaling is known to be weakly violated.\ \cite{schomerus-titov-prb03}

\textcolor{black}{The different definition of the conductance used
here produces results quite similar to the ones obtained with Eq.~(\ref{eq:g-average})
and shown in Fig.~\ref{fig:The-beta-function-2}. In fact, close
inspection of the two sets of curves reveals almost perfect agreement
in the Ohmic and strongly localized regimes but small deviations around
$g=1$.}

\bibliographystyle{apsrev4-1}
\bibliography{referencias}

\begin{thebibliography}{26}%
\makeatletter
\providecommand \@ifxundefined [1]{%
 \@ifx{#1\undefined}
}%
\providecommand \@ifnum [1]{%
 \ifnum #1\expandafter \@firstoftwo
 \else \expandafter \@secondoftwo
 \fi
}%
\providecommand \@ifx [1]{%
 \ifx #1\expandafter \@firstoftwo
 \else \expandafter \@secondoftwo
 \fi
}%
\providecommand \natexlab [1]{#1}%
\providecommand \enquote  [1]{``#1''}%
\providecommand \bibnamefont  [1]{#1}%
\providecommand \bibfnamefont [1]{#1}%
\providecommand \citenamefont [1]{#1}%
\providecommand \href@noop [0]{\@secondoftwo}%
\providecommand \href [0]{\begingroup \@sanitize@url \@href}%
\providecommand \@href[1]{\@@startlink{#1}\@@href}%
\providecommand \@@href[1]{\endgroup#1\@@endlink}%
\providecommand \@sanitize@url [0]{\catcode `\\12\catcode `\$12\catcode
  `\&12\catcode `\#12\catcode `\^12\catcode `\_12\catcode `\%12\relax}%
\providecommand \@@startlink[1]{}%
\providecommand \@@endlink[0]{}%
\providecommand \url  [0]{\begingroup\@sanitize@url \@url }%
\providecommand \@url [1]{\endgroup\@href {#1}{\urlprefix }}%
\providecommand \urlprefix  [0]{URL }%
\providecommand \Eprint [0]{\href }%
\providecommand \doibase [0]{http://dx.doi.org/}%
\providecommand \selectlanguage [0]{\@gobble}%
\providecommand \bibinfo  [0]{\@secondoftwo}%
\providecommand \bibfield  [0]{\@secondoftwo}%
\providecommand \translation [1]{[#1]}%
\providecommand \BibitemOpen [0]{}%
\providecommand \bibitemStop [0]{}%
\providecommand \bibitemNoStop [0]{.\EOS\space}%
\providecommand \EOS [0]{\spacefactor3000\relax}%
\providecommand \BibitemShut  [1]{\csname bibitem#1\endcsname}%
\let\auto@bib@innerbib\@empty
%</preamble>
\bibitem [{\citenamefont {Deych}\ \emph {et~al.}(2000)\citenamefont {Deych},
  \citenamefont {Lisyansky},\ and\ \citenamefont
  {Altshuler}}]{deych-lisyansky-altshuler-prl00}%
  \BibitemOpen
  \bibfield  {author} {\bibinfo {author} {\bibfnamefont {L.~I.}\ \bibnamefont
  {Deych}}, \bibinfo {author} {\bibfnamefont {A.~A.}\ \bibnamefont
  {Lisyansky}}, \ and\ \bibinfo {author} {\bibfnamefont {B.~L.}\ \bibnamefont
  {Altshuler}},\ }\href {\doibase 10.1103/PhysRevLett.84.2678} {\bibfield
  {journal} {\bibinfo  {journal} {Phys. Rev. Lett.}\ }\textbf {\bibinfo
  {volume} {84}},\ \bibinfo {pages} {2678} (\bibinfo {year}
  {2000})}\BibitemShut {NoStop}%
\bibitem [{\citenamefont {Deych}\ \emph {et~al.}(2001)\citenamefont {Deych},
  \citenamefont {Lisyansky},\ and\ \citenamefont
  {Altshuler}}]{deych-lisyansky-altshuler-prb01}%
  \BibitemOpen
  \bibfield  {author} {\bibinfo {author} {\bibfnamefont {L.~I.}\ \bibnamefont
  {Deych}}, \bibinfo {author} {\bibfnamefont {A.~A.}\ \bibnamefont
  {Lisyansky}}, \ and\ \bibinfo {author} {\bibfnamefont {B.~L.}\ \bibnamefont
  {Altshuler}},\ }\href {\doibase 10.1103/PhysRevB.64.224202} {\bibfield
  {journal} {\bibinfo  {journal} {Phys. Rev. B}\ }\textbf {\bibinfo {volume}
  {64}},\ \bibinfo {pages} {224202} (\bibinfo {year} {2001})}\BibitemShut
  {NoStop}%
\bibitem [{\citenamefont {Deych}\ \emph {et~al.}(2003)\citenamefont {Deych},
  \citenamefont {Erementchouk}, \citenamefont {Lisyansky},\ and\ \citenamefont
  {Altshuler}}]{deych-etal-prl03}%
  \BibitemOpen
  \bibfield  {author} {\bibinfo {author} {\bibfnamefont {L.~I.}\ \bibnamefont
  {Deych}}, \bibinfo {author} {\bibfnamefont {M.~V.}\ \bibnamefont
  {Erementchouk}}, \bibinfo {author} {\bibfnamefont {A.~A.}\ \bibnamefont
  {Lisyansky}}, \ and\ \bibinfo {author} {\bibfnamefont {B.~L.}\ \bibnamefont
  {Altshuler}},\ }\href {\doibase 10.1103/PhysRevLett.91.096601} {\bibfield
  {journal} {\bibinfo  {journal} {Phys. Rev. Lett.}\ }\textbf {\bibinfo
  {volume} {91}},\ \bibinfo {pages} {096601} (\bibinfo {year}
  {2003})}\BibitemShut {NoStop}%
\bibitem [{\citenamefont {Abrahams}\ \emph {et~al.}(1979)\citenamefont
  {Abrahams}, \citenamefont {Anderson}, \citenamefont {Licciardello},\ and\
  \citenamefont {Ramakrishnan}}]{abrahams-gang4}%
  \BibitemOpen
  \bibfield  {author} {\bibinfo {author} {\bibfnamefont {E.}~\bibnamefont
  {Abrahams}}, \bibinfo {author} {\bibfnamefont {P.~W.}\ \bibnamefont
  {Anderson}}, \bibinfo {author} {\bibfnamefont {D.~C.}\ \bibnamefont
  {Licciardello}}, \ and\ \bibinfo {author} {\bibfnamefont {T.~V.}\
  \bibnamefont {Ramakrishnan}},\ }\href {\doibase 10.1103/PhysRevLett.42.673}
  {\bibfield  {journal} {\bibinfo  {journal} {Phys. Rev. Lett.}\ }\textbf
  {\bibinfo {volume} {42}},\ \bibinfo {pages} {673} (\bibinfo {year}
  {1979})}\BibitemShut {NoStop}%
\bibitem [{\citenamefont {Anderson}\ \emph {et~al.}(1980)\citenamefont
  {Anderson}, \citenamefont {Thouless}, \citenamefont {Abrahams},\ and\
  \citenamefont {Fisher}}]{anderson-etal-prb80}%
  \BibitemOpen
  \bibfield  {author} {\bibinfo {author} {\bibfnamefont {P.~W.}\ \bibnamefont
  {Anderson}}, \bibinfo {author} {\bibfnamefont {D.~J.}\ \bibnamefont
  {Thouless}}, \bibinfo {author} {\bibfnamefont {E.}~\bibnamefont {Abrahams}},
  \ and\ \bibinfo {author} {\bibfnamefont {D.~S.}\ \bibnamefont {Fisher}},\
  }\href {\doibase 10.1103/PhysRevB.22.3519} {\bibfield  {journal} {\bibinfo
  {journal} {Phys. Rev. B}\ }\textbf {\bibinfo {volume} {22}},\ \bibinfo
  {pages} {3519} (\bibinfo {year} {1980})}\BibitemShut {NoStop}%
\bibitem [{\citenamefont {Cohen}\ \emph {et~al.}(1988)\citenamefont {Cohen},
  \citenamefont {Roth},\ and\ \citenamefont {Shapiro}}]{cohen-etal-prb88}%
  \BibitemOpen
  \bibfield  {author} {\bibinfo {author} {\bibfnamefont {A.}~\bibnamefont
  {Cohen}}, \bibinfo {author} {\bibfnamefont {Y.}~\bibnamefont {Roth}}, \ and\
  \bibinfo {author} {\bibfnamefont {B.}~\bibnamefont {Shapiro}},\ }\href
  {\doibase 10.1103/PhysRevB.38.12125} {\bibfield  {journal} {\bibinfo
  {journal} {Phys. Rev. B}\ }\textbf {\bibinfo {volume} {38}},\ \bibinfo
  {pages} {12125} (\bibinfo {year} {1988})}\BibitemShut {NoStop}%
\bibitem [{\citenamefont {Balents}\ and\ \citenamefont
  {Fisher}(1997)}]{balents-fisher-prb97}%
  \BibitemOpen
  \bibfield  {author} {\bibinfo {author} {\bibfnamefont {L.}~\bibnamefont
  {Balents}}\ and\ \bibinfo {author} {\bibfnamefont {M.~P.~A.}\ \bibnamefont
  {Fisher}},\ }\href {\doibase 10.1103/PhysRevB.56.12970} {\bibfield  {journal}
  {\bibinfo  {journal} {Phys. Rev. B}\ }\textbf {\bibinfo {volume} {56}},\
  \bibinfo {pages} {12970} (\bibinfo {year} {1997})}\BibitemShut {NoStop}%
\bibitem [{\citenamefont {Schomerus}\ and\ \citenamefont
  {Titov}(2003)}]{schomerus-titov-prb03}%
  \BibitemOpen
  \bibfield  {author} {\bibinfo {author} {\bibfnamefont {H.}~\bibnamefont
  {Schomerus}}\ and\ \bibinfo {author} {\bibfnamefont {M.}~\bibnamefont
  {Titov}},\ }\href {\doibase 10.1103/PhysRevB.67.100201} {\bibfield  {journal}
  {\bibinfo  {journal} {Phys. Rev. B}\ }\textbf {\bibinfo {volume} {67}},\
  \bibinfo {pages} {100201} (\bibinfo {year} {2003})}\BibitemShut {NoStop}%
\bibitem [{\citenamefont {Eggarter}\ and\ \citenamefont
  {Riedinger}(1978)}]{eggarter-riedinger-prb78}%
  \BibitemOpen
  \bibfield  {author} {\bibinfo {author} {\bibfnamefont {T.~P.}\ \bibnamefont
  {Eggarter}}\ and\ \bibinfo {author} {\bibfnamefont {R.}~\bibnamefont
  {Riedinger}},\ }\href {\doibase 10.1103/PhysRevB.18.569} {\bibfield
  {journal} {\bibinfo  {journal} {Phys. Rev. B}\ }\textbf {\bibinfo {volume}
  {18}},\ \bibinfo {pages} {569} (\bibinfo {year} {1978})}\BibitemShut
  {NoStop}%
\bibitem [{\citenamefont {Soukoulis}\ and\ \citenamefont
  {Economou}(1981)}]{soukoulis-economou-prb81}%
  \BibitemOpen
  \bibfield  {author} {\bibinfo {author} {\bibfnamefont {C.~M.}\ \bibnamefont
  {Soukoulis}}\ and\ \bibinfo {author} {\bibfnamefont {E.~N.}\ \bibnamefont
  {Economou}},\ }\href {\doibase 10.1103/PhysRevB.24.5698} {\bibfield
  {journal} {\bibinfo  {journal} {Phys. Rev. B}\ }\textbf {\bibinfo {volume}
  {24}},\ \bibinfo {pages} {5698} (\bibinfo {year} {1981})}\BibitemShut
  {NoStop}%
\bibitem [{\citenamefont {Inui}\ \emph {et~al.}(1994)\citenamefont {Inui},
  \citenamefont {Trugman},\ and\ \citenamefont
  {Abrahams}}]{inui-trugman-abrahams-prb94}%
  \BibitemOpen
  \bibfield  {author} {\bibinfo {author} {\bibfnamefont {M.}~\bibnamefont
  {Inui}}, \bibinfo {author} {\bibfnamefont {S.~A.}\ \bibnamefont {Trugman}}, \
  and\ \bibinfo {author} {\bibfnamefont {E.}~\bibnamefont {Abrahams}},\ }\href
  {\doibase 10.1103/PhysRevB.49.3190} {\bibfield  {journal} {\bibinfo
  {journal} {Phys. Rev. B}\ }\textbf {\bibinfo {volume} {49}},\ \bibinfo
  {pages} {3190} (\bibinfo {year} {1994})}\BibitemShut {NoStop}%
\bibitem [{\citenamefont {Fisher}(1995)}]{fisher95}%
  \BibitemOpen
  \bibfield  {author} {\bibinfo {author} {\bibfnamefont {D.~S.}\ \bibnamefont
  {Fisher}},\ }\href {\doibase 10.1103/PhysRevB.51.6411} {\bibfield  {journal}
  {\bibinfo  {journal} {Phys. Rev. B}\ }\textbf {\bibinfo {volume} {51}},\
  \bibinfo {pages} {6411} (\bibinfo {year} {1995})}\BibitemShut {NoStop}%
\bibitem [{\citenamefont {Ma}\ \emph {et~al.}(1979)\citenamefont {Ma},
  \citenamefont {Dasgupta},\ and\ \citenamefont {Hu}}]{MDH-PRL}%
  \BibitemOpen
  \bibfield  {author} {\bibinfo {author} {\bibfnamefont {S.-k.}\ \bibnamefont
  {Ma}}, \bibinfo {author} {\bibfnamefont {C.}~\bibnamefont {Dasgupta}}, \ and\
  \bibinfo {author} {\bibfnamefont {C.-k.}\ \bibnamefont {Hu}},\ }\href
  {\doibase 10.1103/PhysRevLett.43.1434} {\bibfield  {journal} {\bibinfo
  {journal} {Phys. Rev. Lett.}\ }\textbf {\bibinfo {volume} {43}},\ \bibinfo
  {pages} {1434} (\bibinfo {year} {1979})}\BibitemShut {NoStop}%
\bibitem [{\citenamefont {Dasgupta}\ and\ \citenamefont {Ma}(1980)}]{MDH-PRB}%
  \BibitemOpen
  \bibfield  {author} {\bibinfo {author} {\bibfnamefont {C.}~\bibnamefont
  {Dasgupta}}\ and\ \bibinfo {author} {\bibfnamefont {S.-k.}\ \bibnamefont
  {Ma}},\ }\href {\doibase 10.1103/PhysRevB.22.1305} {\bibfield  {journal}
  {\bibinfo  {journal} {Phys. Rev. B}\ }\textbf {\bibinfo {volume} {22}},\
  \bibinfo {pages} {1305} (\bibinfo {year} {1980})}\BibitemShut {NoStop}%
\bibitem [{\citenamefont {Bhatt}\ and\ \citenamefont {Lee}(1982)}]{bhatt-lee}%
  \BibitemOpen
  \bibfield  {author} {\bibinfo {author} {\bibfnamefont {R.~N.}\ \bibnamefont
  {Bhatt}}\ and\ \bibinfo {author} {\bibfnamefont {P.~A.}\ \bibnamefont
  {Lee}},\ }\href {\doibase 10.1103/PhysRevLett.48.344} {\bibfield  {journal}
  {\bibinfo  {journal} {Phys. Rev. Lett.}\ }\textbf {\bibinfo {volume} {48}},\
  \bibinfo {pages} {344} (\bibinfo {year} {1982})}\BibitemShut {NoStop}%
\bibitem [{\citenamefont {Abrahams}(2010)}]{50-years-localization}%
  \BibitemOpen
  \bibinfo {editor} {\bibfnamefont {E.}~\bibnamefont {Abrahams}},\ ed.,\
  \href@noop {} {\emph {\bibinfo {title} {50 {Y}ears of {A}nderson
  {L}ocalization}}}\ (\bibinfo  {publisher} {World Scientific},\ \bibinfo
  {address} {Singapore},\ \bibinfo {year} {2010})\BibitemShut {NoStop}%
\bibitem [{\citenamefont {Johri}\ and\ \citenamefont
  {Bhatt}(2012{\natexlab{a}})}]{johri-bhatt-prl12}%
  \BibitemOpen
  \bibfield  {author} {\bibinfo {author} {\bibfnamefont {S.}~\bibnamefont
  {Johri}}\ and\ \bibinfo {author} {\bibfnamefont {R.~N.}\ \bibnamefont
  {Bhatt}},\ }\href {\doibase 10.1103/PhysRevLett.109.076402} {\bibfield
  {journal} {\bibinfo  {journal} {Phys. Rev. Lett.}\ }\textbf {\bibinfo
  {volume} {109}},\ \bibinfo {pages} {076402} (\bibinfo {year}
  {2012}{\natexlab{a}})}\BibitemShut {NoStop}%
\bibitem [{\citenamefont {Johri}\ and\ \citenamefont
  {Bhatt}(2012{\natexlab{b}})}]{johri-bhatt-prb12}%
  \BibitemOpen
  \bibfield  {author} {\bibinfo {author} {\bibfnamefont {S.}~\bibnamefont
  {Johri}}\ and\ \bibinfo {author} {\bibfnamefont {R.~N.}\ \bibnamefont
  {Bhatt}},\ }\href {\doibase 10.1103/PhysRevB.86.125140} {\bibfield  {journal}
  {\bibinfo  {journal} {Phys. Rev. B}\ }\textbf {\bibinfo {volume} {86}},\
  \bibinfo {pages} {125140} (\bibinfo {year} {2012}{\natexlab{b}})}\BibitemShut
  {NoStop}%
\bibitem [{\citenamefont {Mott}\ and\ \citenamefont
  {Twose}(1961)}]{mott-twose-advphys61}%
  \BibitemOpen
  \bibfield  {author} {\bibinfo {author} {\bibfnamefont {N.~F.}\ \bibnamefont
  {Mott}}\ and\ \bibinfo {author} {\bibfnamefont {W.~D.}\ \bibnamefont
  {Twose}},\ }\href {\doibase 10.1080/00018736100101271} {\bibfield  {journal}
  {\bibinfo  {journal} {Adv. Phys.}\ }\textbf {\bibinfo {volume} {10}},\
  \bibinfo {pages} {107} (\bibinfo {year} {1961})}\BibitemShut {NoStop}%
\bibitem [{\citenamefont {Motrunich}\ \emph {et~al.}(2002)\citenamefont
  {Motrunich}, \citenamefont {Damle},\ and\ \citenamefont
  {Huse}}]{motrunich-damle-huse-prb02}%
  \BibitemOpen
  \bibfield  {author} {\bibinfo {author} {\bibfnamefont {O.}~\bibnamefont
  {Motrunich}}, \bibinfo {author} {\bibfnamefont {K.}~\bibnamefont {Damle}}, \
  and\ \bibinfo {author} {\bibfnamefont {D.~A.}\ \bibnamefont {Huse}},\ }\href
  {\doibase 10.1103/PhysRevB.65.064206} {\bibfield  {journal} {\bibinfo
  {journal} {Phys. Rev. B}\ }\textbf {\bibinfo {volume} {65}},\ \bibinfo
  {pages} {064206} (\bibinfo {year} {2002})}\BibitemShut {NoStop}%
\bibitem [{\citenamefont {M\'elin}\ \emph {et~al.}(2005)\citenamefont
  {M\'elin}, \citenamefont {Dou\ifmmode~\mbox{\c{c}}\else \c{c}\fi{}ot},\ and\
  \citenamefont {Igl\'oi}}]{melin-doucot-igloi-prb05}%
  \BibitemOpen
  \bibfield  {author} {\bibinfo {author} {\bibfnamefont {R.}~\bibnamefont
  {M\'elin}}, \bibinfo {author} {\bibfnamefont {B.}~\bibnamefont
  {Dou\ifmmode~\mbox{\c{c}}\else \c{c}\fi{}ot}}, \ and\ \bibinfo {author}
  {\bibfnamefont {F.}~\bibnamefont {Igl\'oi}},\ }\href {\doibase
  10.1103/PhysRevB.72.024205} {\bibfield  {journal} {\bibinfo  {journal} {Phys.
  Rev. B}\ }\textbf {\bibinfo {volume} {72}},\ \bibinfo {pages} {024205}
  (\bibinfo {year} {2005})}\BibitemShut {NoStop}%
\bibitem [{Note1()}]{Note1}%
  \BibitemOpen
  \bibinfo {note} {Except for extremely singular ones like $P\sim 1/\left
  [\left |t\right |\left |\protect \qopname \relax o{ln}\left |t\right |\right
  |^{x}\right ]$.}\BibitemShut {Stop}%
\bibitem [{\citenamefont {Fisher}\ and\ \citenamefont
  {Young}(1998)}]{fisher-young-RTFIM}%
  \BibitemOpen
  \bibfield  {author} {\bibinfo {author} {\bibfnamefont {D.~S.}\ \bibnamefont
  {Fisher}}\ and\ \bibinfo {author} {\bibfnamefont {A.~P.}\ \bibnamefont
  {Young}},\ }\href {\doibase 10.1103/PhysRevB.58.9131} {\bibfield  {journal}
  {\bibinfo  {journal} {Phys. Rev. B}\ }\textbf {\bibinfo {volume} {58}},\
  \bibinfo {pages} {9131} (\bibinfo {year} {1998})}\BibitemShut {NoStop}%
\bibitem [{\citenamefont {Hoyos}\ \emph {et~al.}(2007)\citenamefont {Hoyos},
  \citenamefont {Vieira}, \citenamefont {Laflorencie},\ and\ \citenamefont
  {Miranda}}]{hoyosvieiralaflorenciemiranda}%
  \BibitemOpen
  \bibfield  {author} {\bibinfo {author} {\bibfnamefont {J.~A.}\ \bibnamefont
  {Hoyos}}, \bibinfo {author} {\bibfnamefont {A.~P.}\ \bibnamefont {Vieira}},
  \bibinfo {author} {\bibfnamefont {N.}~\bibnamefont {Laflorencie}}, \ and\
  \bibinfo {author} {\bibfnamefont {E.}~\bibnamefont {Miranda}},\ }\href
  {\doibase 10.1103/PhysRevB.76.174425} {\bibfield  {journal} {\bibinfo
  {journal} {Phys. Rev. B}\ }\textbf {\bibinfo {volume} {76}},\ \bibinfo
  {pages} {174425} (\bibinfo {year} {2007})}\BibitemShut {NoStop}%
\bibitem [{\citenamefont {Zhou}\ \emph {et~al.}(2009)\citenamefont {Zhou},
  \citenamefont {Hoyos}, \citenamefont {Dobrosavljevi\'c},\ and\ \citenamefont
  {Miranda}}]{zhou-etal-epl09}%
  \BibitemOpen
  \bibfield  {author} {\bibinfo {author} {\bibfnamefont {S.}~\bibnamefont
  {Zhou}}, \bibinfo {author} {\bibfnamefont {J.~A.}\ \bibnamefont {Hoyos}},
  \bibinfo {author} {\bibfnamefont {V.}~\bibnamefont {Dobrosavljevi\'c}}, \
  and\ \bibinfo {author} {\bibfnamefont {E.}~\bibnamefont {Miranda}},\ }\href
  {\doibase 10.1209/0295-5075/87/27003} {\bibfield  {journal} {\bibinfo
  {journal} {Europhys. Lett.}\ }\textbf {\bibinfo {volume} {87}},\ \bibinfo
  {pages} {27003} (\bibinfo {year} {2009})},\ \Eprint
  {http://arxiv.org/abs/arXiv:0810.3043} {arXiv:0810.3043} \BibitemShut
  {NoStop}%
\bibitem [{\citenamefont {Pendry}(1982)}]{pendry-jpc82}%
  \BibitemOpen
  \bibfield  {author} {\bibinfo {author} {\bibfnamefont {J.~B.}\ \bibnamefont
  {Pendry}},\ }\href {http://stacks.iop.org/0022-3719/15/i=28/a=009} {\bibfield
   {journal} {\bibinfo  {journal} {Journal of Physics C: Solid State Physics}\
  }\textbf {\bibinfo {volume} {15}},\ \bibinfo {pages} {5773} (\bibinfo {year}
  {1982})}\BibitemShut {NoStop}%
\end{thebibliography}%

\end{document}